\documentclass[11pt,a4paper]{article}
\pdfoutput=1
\usepackage{jheppub}

\usepackage{subfigure}
\usepackage{amsfonts}
\newcommand{\N}{N_{\bZ}}
\usepackage{pifont}
\usepackage{amsmath}
\usepackage{amssymb}

\usepackage[german,english]{babel} 
\newcommand{\cC}{\mathcal C}

\newcommand{\cN}{\mathcal N}
\newcommand{\cG}{\mathcal G}
\newcommand{\cM}{{\cal M}}
\newcommand{\cO}{{\cal O}}

\newcommand{\eq}[2]{\begin{equation} #1 \label{#2} \end{equation}}

\newcommand{\speq}[2]{\begin{equation} \begin{split} #1 \end{split} \label{#2} \end{equation}}
\newcommand{\al}{\alpha}
\newcommand{\be}{\beta}
\newcommand{\ga}{\gamma}
\newcommand{\de}{\delta}

\newcommand{\la}{\lambda}

\newcommand{\eps}{\epsilon}

\newcommand{\Ga}{\Gamma}

\newcommand{\La}{\Lambda}

\newcommand{\bC}{\mathbb C}

\newcommand{\bR}{\mathbb R}

\newcommand{\del}[1]{\partial_{#1}}

\newcommand{\im}{{\rm Im}\,}
\newcommand{\re}{{\rm Re}\,}

\def\Ouno{\mathcal{O}\left(t_2^{-1}\right)}
\def\Odue{\mathcal{O}\left(t_2^{-2}\right)}
\def\Otre{\mathcal{O}\left(t_2^{-3}\right)}
\def\Oquattro{\mathcal{O}\left(t_2^{-4}\right)}
\def\Ocinque{\mathcal{O}\left(t_2^{-5}\right)}
\def\Osei{\mathcal{O}\left(t_2^{-6}\right)}
\def\unoO{\mathcal{O}\left(t_2\right)}

\def\dueo{o\left(t_2^2\right)}

\def\quattroo{o\left(t_2^4\right)}

\def\seio{o\left(t_2^6\right)}

\def\N{\hat{N}}
\def\Z{\mathbb{Z}}

\title{Restrictions on infinite sequences of type IIB vacua}

\author[a]{Andreas P. Braun,}
\author[a]{Niklas Johansson,}
\author[b]{Magdalena Larfors}
\author[a]{and Nils-Ole Walliser}
\affiliation[a]{Institute for Theoretical Physics, 
           Vienna University of Technology,\\
           Wiedner Hauptstr. 8--10/136,
           A-1040 Vienna, Austria\\ }
\affiliation[b]{Arnold-Sommerfeld-Center for Theoretical Physics\\ Department f\"ur Physik,
Ludwig-Maximilians-Universit\"at M\"unchen\\ Theresienstr. 37, 80333 Munich, Germany}

\emailAdd{abraun@hep.itp.tuwien.ac.at}
\emailAdd{niklasj@hep.itp.tuwien.ac.at}
\emailAdd{magdalena.larfors@physik.uni-muenchen.de}
\emailAdd{walliser@hep.itp.tuwien.ac.at}

\abstract{Ashok and Douglas have shown that infinite sequences of type
IIB flux vacua with imaginary self-dual flux can only occur in
so-called D-limits, corresponding
to singular points in complex structure moduli space. In this work we
refine this no-go result by demonstrating that there are no infinite sequences
accumulating to the large complex structure point of a certain class
of one-parameter Calabi--Yau manifolds. We perform a similar
analysis for conifold points and for the decoupling limit, obtaining identical results. Furthermore, we establish the absence of 
infinite sequences in a D-limit corresponding to 
the large complex structure limit of a two-parameter Calabi--Yau. In particular, our results demonstrate analytically that the series of vacua recently
discovered by Ahlqvist et al.,
seemingly accumulating to the large complex structure point, are
finite. We perform a numerical study of these series close to
the large complex structure point using appropriate approximations for
the period functions. This analysis reveals that the series bounce out
from the large complex structure point, and that the flux eventually
ceases to be imaginary self-dual. Finally, we study D-limits for F-theory
compactifications on $K3\times K3$ for which the finiteness of supersymmetric
vacua is already established. We do find infinite sequences of flux vacua which 
are, however, identified by automorphisms of $K3$.}


\begin{document}

\begin{flushright}
TUW-11-20\\
LMU-ASC 35/11
\end{flushright}

\maketitle

\flushbottom

\section{Introduction}

With our present understanding, string theory seems to allow for a vast number of metastable four-dimensional vacua. This set of universes is often called the string landscape \cite{Susskind:2003kw}, and is equipped with a, in principle computable, effective potential. In a scenario where our universe is described by fluctuations around a particular minimum of this potential, particle masses and couplings are given by local curvatures at the minimum. But there might also be more subtle observational effects depending on the large scale structure of the potential.

One important topographical feature, relevant for many effects in string cosmology, is the existence of sequences of vacua connected by continuous potential barriers. When quantum effects are taken into account, tunnelling can occur between the vacua, with a probability that is computable once the features of the potential barrier are known. From the space-time perspective, the tunnelling process consists of the nucleation of a bubble of the new vacuum inside the old vacuum phase. Depending on the tunnelling rate, and the expansion rates of the new and old universes, the transition to the new vacuum is either complete or partial. The latter case, when part of space-time remains in the old vacuum, is known as eternal inflation \cite{Vilenkin:1983xq,Linde:1986fd,Linde:1986fc}. Potentially, bubble collisions in such a cosmological scenario could leave an observable imprint in the CMBR \cite{Aguirre:2007an, Chang:2007eq}, and this was recently compared with the WMAP 7-year results \cite{Feeney:2010dd,Feeney:2010jj}.

Another interesting possibility is that of chain inflation \cite{Freese:2004vs,Freese:2006fk,Chialva:2008zw,Chialva:2008xh}, see also \cite{Ashoorioon:2010vw}. In this kind of models, inflation results from
sequential tunnelling in a chain of de Sitter universes, each supporting just part of an e-folding. To be a viable option, chain inflation requires the existence of sequences of neighbouring vacua with 
certain properties. Other effects hinging on our local landscape surroundings are resonance tunnelling \cite{HenryTye:2006tg}, ``giant leaps'' between far-away vacua \cite{Brown:2010bc}, and disappearing instantons \cite{Johnson:2008vn,Brown:2011um} --- effects that can greatly affect tunnelling probabilities.
In all these cases, detailed knowledge of the potential is required to obtain quantitative results.

One part of the landscape that offers fairly accurate analytical and numerical control is the complex
structure moduli space of type IIB flux compactifications. Fluxes piercing non-trivial three-cycles of
the internal geometry generate a potential with discrete minima for the complex structure moduli and axio-dilaton. Each of these minima corresponds, after fixing of the 
K\"ahler moduli \cite{Kachru:2003aw, Balasubramanian:2005zx, Conlon:2005ki}, to a vacuum in the landscape. The reason for the mathematical 
tractability of many of these models is that the internal manifold remains conformally Calabi--Yau after the introduction of fluxes, making the powerful tool-kit of special geometry applicable.
Indeed, as demonstrated in Ref. \cite{Taylor:1999ii,Giddings:2001yu}, the potential
of the resulting four-dimensional $\cN = 1$ supergravity is determined by the Gukov--Vafa--Witten superpotential 
\cite{Gukov:1999ya} making it straightforward to compute.

Taking advantage of this fact, type IIB flux compactifications and the resulting potential have been studied in many contexts. In Ref. \cite{Danielsson:2006jg} it was shown that D3-brane black holes, which also affect the potential for the moduli, must be small to affect the minimisation, but can then potentially serve as seeds for bubble nucleation, and in Ref. \cite{Ceresole:2006iq}, explicit profiles of BPS domain walls interpolating between different vacua were obtained. Exploiting the fact that fluxes transform under 
monodromy transformations, it was demonstrated in \cite{Danielsson:2006xw, Chialva:2007sv} that long sequences of continuously connected vacua are a common feature in the landscape, thus opening up the possibility of chain inflation or resonance tunnelling in this framework. These studies were extended in \cite{Johnson:2008kc}, where the tunnelling probabilities between vacua in the sequence were first computed. Moreover, Ref. \cite{Johnson:2008vn,Aguirre:2009tp} investigated the influence of the universal K\"ahler modulus on tunnelling rates and domain wall dynamics in this setting.

Recently, Ahlqvist et al. \cite{Ahlqvist:2010ki} continued these investigations of type IIB vacuum sequences by a thorough study of a class of one-parameter Calabi--Yau models. This revealed several intriguing features, both in the tunnelling dynamics 
and in the vacuum structure. For vacua connected by conifold monodromies, it was demonstrated that the 
tunnelling trajectories tend to pass close to the conifold point. Furthermore, long sequences of connected minima
seemingly accumulating to the large complex structure (LCS) point were found. It was left as an open question whether these
sequences continue indefinitely or not. That the sequences approach the LCS point is particularly interesting in view
of the no-go theorem \cite{Ashok:2003gk} derived by Ashok and Douglas stating that infinite sequences of vacua 
with imaginary self-dual (ISD) flux can only occur if they accumulate to so-called D-limits --- one example of which is 
the LCS point. It is the aim of the present paper to investigate if it is possible to have infinite sequences of minima
accumulating to the LCS point, and in particular to determine whether the sequences found in \cite{Ahlqvist:2010ki} 
end or not. 

Previous studies \cite{Denef:2004ze,Acharya:2006zw, Eguchi:2005eh, Torroba:2006kt} of the finiteness of type IIB flux vacua have mainly been based on the statistical methods pioneered in 
\cite{Bousso:2000xa, Douglas:2003um}. (For reviews, consult e.g. \cite{Douglas:2006es, Denef:2008wq}.) This approach uses a continuum approximation of the fluxes, which allows to relate
the density of vacua in moduli space to the Euler density of a certain metric on moduli space. As this Euler density is an index, it only gives a lower bound for the true vacuum density in principle. 
The index can be shown to integrate to finite values around regular points in moduli space \cite{Ashok:2003gk}. Its structure around Calabi--Yau singularities has been analysed in \cite{Eguchi:2005eh, Torroba:2006kt}, 
including both ADE singularities and the LCS points. In both cases the index integrates to finite results. For results concerning the finiteness of the intersecting D-brane landscape, see
e.g. \cite{Blumenhagen:2004xx,Gmeiner:2005vz,Douglas:2006xy}.

In this work we complement these statistical studies with a more direct analysis of the possibility of having any infinite sequence of
ISD vacua. By analysing the geometry of the complex structure moduli space we flesh out the details of the Ashok--Douglas theorem and 
obtain explicit expressions for a positive definite quadratic form that must stay finite in any sequence of ISD vacua.
Using this, we derive an extension of the no-go theorem. Namely, for one-parameter Calabi--Yau manifolds, there are no
sequences of vacua accumulating to the LCS point. This shows by analytical means that, in particular, the sequences
of \cite{Ahlqvist:2010ki} are finite. We furthermore extend this result to the D-limits corresponding to conifold points
and decoupling limits, and also study the LCS limit of a two-parameter model. 

In addition, we treat the case when the compactification manifold is $K3 \times K3$, for which the finiteness is proven
in a very different manner \cite{Aspinwall:2005ad}. We find that D-limits exist and that infinite sequences can be constructed.
Hence all but a finite number of the solutions in a sequence must be related by automorphisms of $K3$. We demonstrate this in
a simple example. Finally, we use numerical methods to study two particular examples of one-parameter Calabi--Yau manifolds. Using
expansions of the periods in the LCS region allows us to efficiently compute the scalar potential, and thus follow the sequences of \cite{Ahlqvist:2010ki} closer to the LCS point.
In accordance with the general analysis, the minima eventually leave the region close to the LCS point.

This paper is organised as follows. Section \ref{sect:IIBmod} contains a review on type IIB compactifications and introduces our notation and conventions. We then discuss the no-go theorem by Ashok and Douglas and the relevance of D-limits in section \ref{sec:dlimits}. Subsequently, we analyse the length of sequences of ISD vacua in various D-limits in type IIB compactifications. Details of this computation is relegated to appendix \ref{app1}. In section \ref{sec:dlimK3}, we discuss sequences of vacua in D-limits in F-theory compactifications on $K3 \times K3$. Through a numerical analysis, we then map out two sequences of type IIB vacua in the LCS region in section \ref{sec:nummod}. Finally, we summarise and discuss our results.

\section{Type IIB moduli stabilisation}\label{sect:IIBmod}

In this section we give a brief review of moduli stabilisation in type IIB supergravity, and set the notation and conventions.
To aid comparison with that work we use as far as possible the notation of \cite{Ahlqvist:2010ki}.

\subsection{Calabi--Yau geometry}

We denote by $\cM$ a Calabi--Yau manifold with complex structure moduli space $M$, and let $\cC$ be the combined moduli space of
complex structure and the axio-dilaton: $(z,\tau)\in \cC$. The periods of
${\cal M}$ are 
\eq{
\Pi_{I}=\int_{C_{I}}\Omega=\int_{\cM}C_{I}\wedge\Omega\, ,
}{eq:2b1}
where $\Omega$ is the holomorphic three-form and $C_{I}$ is a basis in
$H_{3}(\cM)$. Note that $C_I$ is used to denote two things: both the cycles and their Poincar\'{e} duals. 
The intersection matrix $Q = (Q_{IJ})$ is defined by
\eq{
Q_{IJ}=\int_{C_{I}}C_{J}=\int_{\cM}C_{I}\wedge C_{J}.
}{eq:2b2}
The periods are collected into a vector whose entries we number in reverse order
\eq{
\Pi(z)=
\begin{pmatrix}
\Pi_{N}(z)\\
\Pi_{N-1}(z)\\
\vdots\\
\Pi_{0}(z)
\end{pmatrix},
}{eq:2b3}
where $z$ is a $h^{(1,2)}$-dimensional (complex) coordinate on $\mathcal{M}$ and $N \equiv 2h^{1,2} + 1$.

In our one-parameter examples there are three special points in moduli space: the large complex structure (LCS) point, the conifold point and the Landau--Ginzburg point. We fix these to lie at $z=0,1$ and $z= \infty$, respectively.
The periods are subject to monodromies upon transport around these points:
\eq{
\Pi\rightarrow T \cdot\Pi,
}{eq:2b4}
where $T$ is a matrix that preserves the symplectic structure $Q$. The complex structure moduli space is furthermore equipped with a K\"ahler metric, with 
K\"ahler potential
\eq{
K_{\mathrm{cs}} = -\log \left( i\int \limits_{\cM} \Omega \wedge \bar{\Omega} \right) = -\log(i\Pi^{\dagger}\cdot Q^{-1} \cdot \Pi) \, .
}{eq:2b9}
Note that our integration conventions are such that 
\eq{
\int_{\cM} \bar{\Ga} \wedge *\Ga > 0
}{eq:2b11}
for any non-zero three-form $\Ga$. Finally, we define \cite{Denef:2000nb} the (antisymmetric, topological, moduli independent) intersection 
product and the (symmetric, positive definite, moduli dependent) scalar product as
\begin{align}
&\langle A_{(3)}, B_{(3)} \rangle  = \int_{\cM} A_{(3)} \wedge B_{(3)} = A \cdot Q \cdot B^T \\
&(A_{(3)}, B_{(3)}) = \int_{\cM} A_{(3)} \wedge *B_{(3)} = A \cdot \cG_z \cdot B^T,
\end{align}
respectively. Here, $A = (A^0, \ldots A^N)$ is a row vector collecting the components of the form $A_{(3)}$ in the basis $C_I$: $A_{(3)} = -A^I C_I$, and similarly for $B$. The matrix $\cG_z$ is a moduli dependent positive quadratic form on $\bC^{N+1}$.

\subsection{Flux vacua}

Fluxes piercing the three-cycles induce a Gukov--Vafa--Witten superpotential $W$, leading to an ${\cal N}=1$ scalar potential
potentially stabilising all complex structure moduli and the axio-dilaton. The potential is
\eq{
V\left(z,\tau\right)  =e^{K}\left(  g^{i\bar{\jmath}}D_{i}WD_{\bar{\jmath}%
}\bar{W}+g^{\tau\bar{\tau}}D_{\tau}WD_{\bar{\tau}}\bar{W}+g^{\rho\bar{\rho}%
}D_{\rho}WD_{\bar{\rho}}\bar{W}-3|W|^{2}\right)\, ,
}{eq:2b5}
where $g^{A\bar{B}} = (\partial_{A}\partial_{\bar{B}}K)^{-1}$ and 
$D_A = \partial_A + \partial_A K$ with $K$ being the ${\cal N}=1$ K\"{a}hler potential. 
To compute $V$ all that is needed are expressions for the superpotential $W$ and the K\"{a}hler 
potential $K$. We denote the three-form fluxes by $F_{(3)}$ (RR) and $H_{(3)}$ (NSNS). We collect the flux quanta in row vectors $F = (F^0, \ldots, F^N)$ defined by
\eq{
F^I = -(Q^{-1})^{IJ}\int_{C_J} F_{(3)}\, ,
}{eq:2b6.5}
and similarly for $H$. Note that the vectors $F$ and $H$ are subject to Dirac quantisation. Their entries are integer multiples of
$4\pi^2 \al'$ which we fix to unity for convenience. For notational convenience we often use the combined three-form flux
\eq{
G_{(3)} = F_{(3)} - \tau H_{(3)}
}{eq:2b9.5}
which can also be represented by a vector, albeit with non-integer components
\eq{
G = F - \tau H\, .
}{eq:2b9.6}
The superpotential is given by
\eq{
W=\int \limits_{\cM}\Omega\wedge G_{(3)} = G \cdot \Pi \,.
}{eq:2b7}
The K\"{a}hler potential is 
\eq{
K=-\ln\left( -i(\tau-\bar{\tau})\right)  + K_{\mathrm{cs}}\left(z,\bar
{z}\right)  -3\ln\left(  -i(\rho-\bar{\rho})\right),
}{eq:2b8}
where $K_{\mathrm{cs}}$ is the K\"{a}hler potential on complex structure moduli space, given by \eqref{eq:2b9}.
Due to the last term in the K\"ahler potential, the contributions of $g^{\rho\bar{\rho}%
}D_{\rho}WD_{\bar{\rho}}\bar{W}$ and $-3|W|^{2}$ cancel:
\eq{
V\left(z,\tau\right)  =e^{K}\left(  g^{i\bar{\jmath}}D_{i}WD_{\bar{\jmath}%
}\bar{W}+g^{\tau\bar{\tau}}D_{\tau}WD_{\bar{\tau}}\bar{W} \right). %
}{eq:2b6}
Using (\ref{eq:2b7}), (\ref{eq:2b8}) and (\ref{eq:2b9}), the scalar potential can be computed numerically once the periods and their 
derivatives are known.

The three-form fluxes induce a D3-brane charge density, that must be compensated by localised sources on the compact manifold.
This amounts to the tadpole condition
\eq{
\int_{\cM} F_{(3)} \wedge H_{(3)} = \frac{\chi}{24}+\tfrac{1}{4}N_{\rm O3} - N_{\rm D3}\equiv L\, ,
}{eq:2b0}
where $N_{\rm O3}$ is the number of O3 planes, $N_{\rm D3}$ is the number of (space filling) D3 branes in the compactification
and $\chi$ counts the tadpole contribution of D7 branes and O7-planes. From the F-theory perspective, $\chi$ is the Euler chracteristic of 
(an appropriate resolution of) the corresponding elliptic Calabi-Yau four-fold.
Expressed in the vectors $F$ and $H$ the tadpole condition reads
\eq{
F \cdot Q \cdot H^T = L.
}{eq:2b10}
Giddings, Kachru and Polchinski \cite{Giddings:2001yu} showed how to solve all equations of motion in the above set-up (see also \cite{Taylor:1999ii}).
For sources satisfying a certain ``BPS-like" condition and at tree level, the equations forces the flux to be imaginary self dual (ISD):
$*G_{(3)} = iG_{(3)}$. This is a condition on the complex structure moduli and the axio-dilaton.
In fact, the ISD condition is equivalent to the vanishing of the F-terms related to these moduli. 
We have the equivalences
\speq{
{\rm EOMs} \iff *G_{(3)} = iG_{(3)} &\iff G_{(3)} \in H^{(2,1)}(\cM)\oplus H^{(0,3)}(\cM)\\
D_{\tau} W = 0 &\iff G_{(3)}^{(3,0)} = 0 \\
D_{i} W = 0 &\iff G_{(3)}^{(1,2)} = 0 \\
D_{\rho} W = 0 \iff W = 0 &\iff G_{(3)}^{(0,3)}=0 
}{eq:2b0.2}
where the harmonic representative is understood by the $\in$ in the first line. We see that ISD implies that
$D_{\tau} W =D_{i} W = 0$ but that supersymmetry can well be broken by a non-zero $D_{\rho} W$ in ISD minima.

Note furthermore that if  $z$ and $\tau$ are tuned so that the flux is ISD, then the potential \eqref{eq:2b6}
has a global minimum. These minima are discrete, and each such configuration corresponds to an ISD vacuum of the
type IIB landscape.

\section{Series in D-limits}
\label{sec:dlimits}

Let us now turn to the question of the existence of infinite sequences of ISD vacua. Ashok and Douglas \cite{Ashok:2003gk} have formulated a no-go theorem that restricts the possibility of infinite sequences of vacua. They also demonstrated that this theorem can be evaded in the vicinity of special points  --- so-called D-limits --- in the moduli 
space $\cC$, one example being the point of large complex structure. In this section we review and make this theorem and the concept of D-limits
more precise.

\subsection{The no-go theorem of Ashok and Douglas}

The two main ingredients of the argument are the tadpole and the ISD conditions. Suppose that we have no anti-D3-branes, and that 
the flux $G_{(3)}$ is imaginary self dual. These conditions include all supersymmetric vacua, but as explained above also other minima. We shall keep 
these two assumptions throughout this section. We then have that
\eq{
\langle F_{(3)},  H_{(3)}\rangle = \frac{\chi}{24}+\tfrac{1}{4}N_{\rm O3}-N_{D3} = L 
}{eq:dlim1}
with $L$ being a number bounded from above since $N_{D3} \geq 0$. On the other hand a short computation yields
\eq{
\langle F_{(3)},  H_{(3)}\rangle = \frac{i}{2\, \im{\tau}}\langle \bar{G}_{(3)} , G_{(3)} \rangle = \frac{1}{2\, \im{\tau}}\langle 
\bar{G}_{(3)}, *G_{(3)}\rangle = \frac{1}{2\, \im{\tau}}\bar{G} \cdot \cG_z \cdot G^T > 0
}{eq:dlim2}
where, in the second step, the imaginary self-duality of $G_{(3)}$ was used. So, in fact, for our type of vacua we have
\eq{
\frac{1}{2\, \im{\tau}} \bar{G} \cdot \cG_z \cdot G^T  = L \leq L_{\rm max} = \frac{\chi}{24}+\tfrac{1}{4}N_{\rm O3}.
}{eq:dlim3}
Since $\cG_z$ is a positive quadratic form we have thus shown that $G$ must lie inside a moduli dependent ellipsoid in $\bC^{N+1}$.
 Let us use this to derive a restriction on the integer valued vectors $F$ and $H$. If we collect these into a $(2N + 2)$-dimensional 
vector $\N = (F,H)$ we have
\eq{
\langle F_{(3)},  H_{(3)}\rangle = \frac{1}{2\, \im{\tau}}\bar{G} \cdot \cG_z \cdot G^T = \N \cdot (\cG_\tau \otimes \cG_z) \cdot \N^T
}{eq:dlim4}
where $\cG_\tau$ is proportional to the metric on a torus with complex structure $\tau$:
\eq{
\cG_\tau = \frac{1}{2\im \tau}\begin{pmatrix}1 & -\re \tau \\ -\re \tau & |\tau|^2 \end{pmatrix}.
}{eq:dlim5}
Thus, the integer vector $\hat{N}$ must lie within an ellipsoid in $\bR^{2N+2}$ whose dimensions are given by 
the $(z,\tau)$-dependent eigenvalues $\La_i$ of the matrix $\cG_\tau \otimes \cG_z$. It is now simple to formulate the no-go result 
of \cite{Ashok:2003gk}.
Any region $R \subset \cC$ of $(\tau, z)$-space for which the $\La_i$ are bounded from below by some positive number, can support 
only a finite number of vacua. To see this, suppose that $\La_i(z,\tau) > \eps$ for all $(z,\tau)$ and $i$. 
Then all admissible $\N$ lie within a ball of radius squared $r^2 = L_{\rm max}/\eps$. These are of course finitely many.

It is also immediately clear how to evade this no-go result. Infinite series of vacua can occur only if their location in $\cC$ 
approaches a point where the matrix $\cG_\tau \otimes \cG_z$ develops a null eigenvector. Points where this happens are referred to as D-limits.

\subsection{D-limits}

Since the eigenvalues of a product matrix is the product of the eigenvalues of the factors, a D-limit can arise in two ways.
Either $\cG_\tau$ or $\cG_z$ can degenerate. In the first case, using S-duality to restrict $\tau$ 
to lie in the standard fundamental domain of the torus moduli space, the only locus where $\cG_\tau$ degenerates is as $\im\tau \to \infty$. This limit 
corresponds to a decoupling limit, and the null eigenvector has only RR-flux.

The other option is that $\cG_z$ degenerates. To find out when this happens we need to compute this matrix in terms of the periods.
Using the expression for $(A_{(3)}, B_{(3)})$ given in Eq. (2.18) of \cite{Denef:2000nb} some simple algebra yields
\eq{
A \cdot \cG_z \cdot B^T = 2 e^K \re \left[ (A\cdot\Pi) (\Pi^{\dagger} \cdot B^T) + g^{i\bar{\jmath}} (A\cdot D_i\Pi) (\bar{D}_{\bar{\jmath}} \Pi^{\dagger} \cdot B^T) \right]
}{eq:dlim7}
so that
\eq{
\cG_z = 2 e^K \re \left[\Pi \,  \Pi^\dagger + g^{i\bar{\jmath}}D_i\Pi \, \bar{D}_{\bar{\jmath}} \Pi^\dagger \right].
}{eq:dlim8}
This matrix can be computed straightforwardly when the periods $\Pi_I$ are known. In section \ref{sect:IIBseries} we shall do this for the large complex structure and conifold limits.

\subsection{D-limits and F-theory}\label{Fselfduality}

Flux compactifications of Type IIB string theory can be embedded in the more general framework of F-theory compactified on elliptic Calabi-Yau fourfolds, see e.g. \cite{Denef:2008wq}. 
On the one hand, F-theory geometrizes the $SL(2,\mathbb{Z})$ self-duality of type IIB string theory. For generic points in moduli space, F-theory models have no interpretation 
in terms of perturbative type IIB string theory due to the presence of various types of $(p,q)$-branes. In Sen's weak coupling limit \cite{Sen:1997kw}, however, F-theory 
reduces to weakly coupled type IIB string theory compactified on Calabi-Yau orientifolds with O7-planes and D7-branes. In F-theory, the closed string moduli are unified with the open string
moduli in the moduli space of the elliptic Calabi-Yau manifold. On the other hand, F-theory can be obtained as a limit of M-theory compactifications on elliptic Calabi-Yau manifolds by 
collapsing the elliptic fibre. As M-theory contains a four-form field strenght, one can introduce four-form fluxes $G_{(4)}$. These must have one leg in the elliptic fibre in order not 
to spoil Lorentz invariance \cite{Dasgupta:1999ss}. In Sen's weak coupling limit, the four-form fluxes $G_{(4)}$ on the M-theory side encode both the three-form flux $G_{(3)}$ as well 
as (abelian) two-form fluxes $F_{(2)}$ on D7-branes on the type IIB side.

The analysis of the last section can be carried over to this case:  
In the absence of $O3$-planes, the condition for the cancellation of the $D3$-brane tadpole is 
\begin{equation}\label{Mtadpole}
\frac{\chi(CY_4)}{24}- \frac{1}{2}\int_{CY_4}G_{(4)}\wedge G_{(4)}=N_{D3} \, .
\end{equation}
Here, $\chi(CY_4)$ denotes the Euler characteristic of the elliptic Calabi--Yau fourfold.

As shown in \cite{Becker:2001pm}, the equations of motion enforce that
\begin{equation}
 \ast G_{(4)}=G_{(4)} \, ,
\end{equation}
 so that
\begin{equation}
 \frac{1}{2}\int_{CY_4}G_{(4)}\wedge G_{(4)} = \frac{1}{2}\int_{CY_4}G_{(4)}\wedge\ast G_{(4)} \geq 0 \, .
\end{equation}
As before, infinite sequences of flux vacua can only exist in a limit in which this positive definite
form develops a zero eigenvector.

\section{Series in type IIB D-limits}\label{sect:IIBseries}

In this section we analyse the possibility of infinite sequences in various D-limits in type IIB compactifications. We treat in turn
the large complex structure limit, decoupling limits and the conifold limit. We assume all the time that only one of these 
special loci is approached, i.e., we do not treat a simultaneous decoupling and LCS limit. In all cases we find that no infinite sequences of ISD vacua are possible.

\subsection{Series around a large complex structure point}\label{sect:lcs}

An example of a D-limit that is ubiquitous in Calabi--Yau moduli spaces is the large complex structure point. 
Since the series of \cite{Ahlqvist:2010ki} seem to accumulate at this point it is natural to investigate whether
such series can continue indefinitely or not. We study therefore one-parameter models with an LCS point and use the no-go
results of Ashok and Douglas. The complex
structure modulus is conventionally denoted by $t = t_1 + it_2$ with $t_{1,2}\in \bR$ and the LCS point is at $t_2 \to \infty$.
For a one-parameter model, the period vector takes the following general form around the LCS point
\eq{
\begin{pmatrix} \Pi_3 \\ \Pi_2 \\ \Pi_1 \\ \Pi_0  \end{pmatrix} \sim \begin{pmatrix} \al_3 \, t^3 + \ga_3 \, t + i\de_3 
\\ \be_2 \, t^2 + \ga_2 \, t + \de_2 \\ t \\ 1
\end{pmatrix}.
}{eq:LCS12} 
Using Eqs. \eqref{eq:dlim8}, \eqref{eq:2b9}, the definition $g_{i\bar\jmath} = \del i \del{\bar\jmath} K$, and the expansion of the periods now allows for a straightforward computation of $\cG_t$. The computation is outlined in Appendix \ref{app1},  and a generic\footnote{We assume that two of the expansion coefficients in \eqref{eq:LCS12} are related as $\be_2 = 3\al_3$. 
This is true for all models in \cite{Ahlqvist:2010ki} and seems to be a general feature. Treating the case $\be_2 \neq 3\al_3$ produces results identical to those presented here.} model of our type gives the result
\eq{
\cG_t = \begin{pmatrix} a_{11} \, t_2^3 + \cO(t_2) & a_{12} \, t_2 + \cO(1/t_2) & a_{13}\, \frac{1}{t_2} + \cO(1/t_2^3) & a_{14}\frac{1}{t_2^3} + \cO(1/t_2^5) \\
\cdot & a_{22} t_2 + \cO(1/t_2) & a_{23} \frac{1}{t_2} + \cO(1/t_2^3) & a_{24}\frac{1}{t_2^3} + \cO(1/t_2^5)\\
\cdot & \cdot & a_{33}\frac{1}{t_2} + \cO(1/t_2^3) & a_{34}\frac{1}{t_2^3} + \cO(1/t_2^5)\\
\cdot &\cdot & \cdot  & a_{44}\frac{1}{t_2^3} + \cO(1/t_2^5)
        \end{pmatrix}
}{eq:LCS13}
for some known constants $a_{ij}$. Here, the entries $\cdot$ in $\cG_t$ are determined by symmetry.  It is clear that this matrix develops two null eigenvectors as $t_2 \to \infty$.

We now prove that there are no infinite series of ISD vacua accumulating at the complex structure point
for our one-parameter models.
Let us begin by noting that the intersection matrix in the basis of \eqref{eq:LCS12} is anti-diagonal:
\eq{
Q_{03} = -Q_{12} = -1 \, .
}{eq:LCS14}
Therefore a flux configuration with $F^0 = F^1 = H^0 = H^1= 0$ satisfies
\eq{
\langle F_{(3)},  H_{(3)} \rangle = 0\, , 
}{eq:LCS15}
implying that for any ISD vacuum corresponding to such fluxes
\eq{
\N \cdot (\cG_\tau \otimes \cG_t) \cdot \N^T = 0 \,.
}{eq:LCS16}
Since the matrix $\cG_t$ is positive definite for any smooth manifold, \eqref{eq:LCS16} implies that the compactification manifold is singular, i.e., that the vacuum sits exactly at the D-limit. In fact, as remarked in \cite{Danielsson:2006xw}, the flux potential always has a minimum at the LCS point for such flux configurations. 
What we shall demonstrate below is that this is the only possibility: there are no series for which one of $F^{0,1}$, $H^{0,1}$ is nonzero.

Since we assume that $\im \tau$ stays finite, the essential features can be deduced from the structure of $\cG_{t}$. We prove first the following
statement. Suppose $\{N_n = (N^0_n, N^1_n, N^2_n, N^3_n)\}$ is a series of integer four-vectors and that $\{t_n\}$ is a series of points in complex structure moduli space such that
\eq{
\lim_{n\to \infty} \im t_n = \infty \qquad \lim_{n\to \infty} N_n \cdot \cG_{t_n} \cdot N_n^T \equiv 
\lim_{n\to \infty} \|N_n\|_t^2 \neq \infty \, . 
}{eq:LCS17}
Then $N_n^0 = N_n^1 = 0$ for $n$ sufficiently large. We prove this by contradiction. To reduce clutter, let us from now on suppress the subscript on $t$ and $N$. Suppose first that $N^0$ is non-zero, without loss of generality let $N^0 = 1$
and assume that \eqref{eq:LCS17} holds.
Denote the eigenvectors and (positive) eigenvalues of the matrix \eqref{eq:LCS13} by $w_i$ and $\la_i$, respectively.
The scalar product $\|N\|_t^2$ can be expanded in this eigenbasis:
\begin{equation}\label{eqn:summands}
\|N \|_t^2 = \sum_{i=1}^{4} \,| N \cdot w_i(t) |^2\lambda_i(t)\, .
\end{equation} 
Since all of the terms in this expression are positive, all of them must stay finite in the limit $t_2\to \infty$.
Expanding in $t_2$, the eigenvalues and eigenvectors are given by
\begin{align}\label{eqn:eigen}
&\lambda_1 = a_{11} \,t_2^3 +\unoO\,,		 &w_1^T =& \left[1, \Odue, \Oquattro , \Osei	\right]\, , \nonumber\\
&\lambda_2 = a_{22}\,t_2  +\Ouno\,,		&w_2^T =& \left[\Odue, 1, \Odue , \Oquattro 	\right]\, , \nonumber\\
&\lambda_3 = \frac{a_{33}}{t_2}+\Odue\,,	&w_3^T =& \left[\Oquattro, \Odue, 1 , \Odue	\right]\, , \nonumber\\
&\lambda_4 = \frac{a_{44}}{t_2^3}+\Ocinque\,,  &w_4^T =& \left[\Osei, \Oquattro, \Odue , 1 \right] \, .
\end{align}
The first eigenvalue grows as $\la_1\sim t_2^3$. Therefore we must have $|N \cdot w_1|^2 \sim\Otre$. Hence
\eq{
\mathcal{O}\left(t_2^{-3/2}\right)= N \cdot w_1 
= 1 + N^1 \Odue + N^2 \Oquattro + N^3\Osei \,.
}{eq:LCS18}
This can happen only if at least one of the $N^i$ diverges. It is also clear that this must happen in order for $N$ to approach one
of the zero eigenvectors of $\cG_t$. What is needed is 
\eq{
N^1 = P \,t_2^2 + \dueo\, , \qquad N^2 = Q\, t_2^4+\quattroo\, , \qquad N^3= R\, t_2^6 +\seio
}{eq:LCS18.5}
where, e.g., $o(t_2^2)$ denotes terms that grows slower than $t_2^2$ and $P$, $Q$ and $R$ are appropriately chosen constants.
Consider now the term in \eqref{eqn:summands} proportional to $\la_4$. We obtain
\eq{
N \cdot w_4  =\Osei+ N^1\Oquattro+ N^2 \Odue+ N^3 \sim R\, t_2^6
}{eq:LCS19}
Hence $| N \cdot w_4 |^2  \la_4 \sim R^2 \, t_2^{9}$, and $R$ must vanish. Considering now in order the terms proportional to
$\la_3$ and $\la_2$ demonstrates in complete parallel that also $Q$ and $P$ must be zero. This is, however, incompatible with \eqref{eq:LCS18},
and we have reached a contradiction. We have thus proved that $N^0 = 0$. Assuming now a flux of the form $N = (0, 1, N^2, N^3)$ and going through an almost identical argument demonstrates $N^1 = 0$.

To complete the argument we now consider a series of integer eight-vector $\N_n = (F_n, H_n)$ and assume that 
\eq{
\N \cdot \cG_\tau \otimes \cG_t \cdot \N^T
}{eq:LCS20}
stays finite as the LCS point is approached. (Again we suppress the index on $\N$ and $t$.) We furthermore assume that $\tau = \tau_1 + i\tau_2$ lies in the standard fundamental domain $|\tau|\geq 1$, $|\tau_1| \leq 1/2$. Since we, by assumption, do not approach a decoupling limit, the whole series fulfils $\tau_2 \leq M$ for some number $M$. This means that the eigenvalues $\mu_{1,2}$ of the matrix \eqref{eq:dlim5} are bounded from below. The eigenvalues and orthonormal eigenvectors $v_{1,2}$ are
\begin{equation}\label{eqn:eigenvalueGT}
\mu_{1,2} = \frac{1}{4\tau_2}\left[1+|\tau|^2 \pm \sqrt{\left(1-|\tau|^2\right)^2+4\tau_1^2}\right] \, , \qquad v_{1,2}  = 
	\begin{pmatrix}
	v_{1,2}^F \\ v_{1,2}^H 
	\end{pmatrix} \, .
\end{equation}
The only property of $v_{1,2}$ important to us presently is their orthonormality. To see that the eigenvalues are bounded note that
\eq{
\mu_1 \geq \mu_2  \geq \frac{1}{4\, \tau_2} \left(2-\sqrt{4\tau_1^2}\right) = \frac{1}{2\, \tau_2} \left(1-\tau_1 \right) > \frac{1}{2\,M} \left(1-\tau_1 \right) \geq \frac{1}{4M} \, .
}{eq:LCS21}
We can now expand in eigenvectors
\eq{
\N \cdot \cG_\tau \otimes \cG_t \cdot \N^T = \sum_{i,j} | \N \cdot v_i\otimes w_j |^2\mu_i \la_j=\sum_{i,j} |\eps_{ij}|^2\la_j\mu_i
}{eq:LCS22}
where 
\eq{
\eps_{ij}\equiv \N \cdot v_i\otimes w_j =v_i^F\left(F\cdot w_j\right)+v_i^H\left(H\cdot w_j\right)\,.
}{eq:LCS23}
Again, each term in the sum \eqref{eq:LCS22} has to stay finite in the limit. Since the $\mu_i$ are bounded, the quantities
$\eps_{1j}$ and $\eps_{2j}$ must each satisfy 
\eq{
\eps_{ij} = \cO(1/\sqrt{\la_j})\, .
}{eq:LCS24}
Using the orthonormality of $v_{1,2}$ \eqref{eq:LCS23} is easily inverted.
In matrix notation
\eq{
	\begin{pmatrix}
	F \cdot w_j \\ H\cdot w_j
	\end{pmatrix} = 	
\begin{pmatrix}
	v_1^F && v_2^F	\\
	v_1^H && v_2^H		
	\end{pmatrix}
	\begin{pmatrix}
	\eps_{1j} \\ \eps_{2j}
	\end{pmatrix}	\, .
}{eq:LCS25}
Since the $v^{F/H}_i$ are bounded non-zero numbers we have therefore proven that 
\eq{
F \cdot w_j = \cO(1/\sqrt{\la_j})\, , \qquad H \cdot w_j = \cO(1/\sqrt{\la_j})\,.
}{eq:LCS26}
This is exactly what is needed to prove that $F^0 = F^1 = H^0 = H^1 = 0$ from the
structure of $\cG_t$, starting from Eq. \eqref{eq:LCS18}.

To sum up, requiring the finiteness of $\N \cdot \cG_\tau \otimes \cG_t \cdot \N^T $ in the limit $t_2 \to \infty$
implies that $F^0 = F^1 = H^0 = H^1 = 0$. This in turn implies that there is no vacuum, except the singular one located exactly 
at the LCS point.

\subsection{Series in decoupling limits}

Let us now, in a very similar manner, prove that there can be no sequences of ISD vacua converging to a decoupling limit.
We consider some flux compactification on a Calabi--Yau whose matrix $\cG_z$ has eigenvalues and vectors $\la_j$ and $w_j$.
(Note that, in this subsection, we do not make any assumptions concerning the dimensionality $b_3$ of the vectors $w_j$, $F$ and $H$.)

Any sequence of ISD vacua must, of course, still have a finite constant value for the quantity in Eq. \eqref{eq:LCS22}, 
and each of the terms in that equation must thus be finite. 
This time however, we assume that the eigenvalues $\la_j$ are bounded from below, and that $\tau_2 \to \infty$. 
The quantities $\eps_{ij}$ therefore must satisfy
\eq{
\eps_{ij} = (v_i^F F + v_i^H H) \cdot  w_j = \cO(1/\sqrt{\mu_i}) \, .
}{eq:dec1}
Using the fact that the $w_j$, as eigenvectors of a symmetric matrix, are orthonormal, it is possible to invert the above relation to yield
\eq{
v_i^F F + v_i^H H = \cO(1/\sqrt{\mu_i})\, .
}{eq;dec2}
In the decoupling limit the eigenvectors $v_i$ and eigenvalues $\mu_i$ are given by
\begin{equation}\label{eq:evGt_exp}
\mu_{1} =  \frac{\tau_2}{2} + \frac{\tau_1^2}{\tau_2} + \ldots\, , \qquad \mu_{2} =  \frac{1}{2\tau_2} - \frac{\tau_1^2}{\tau_2^3} + \ldots \, ,
\end{equation}
\eq{
v_1 =\begin{pmatrix}
	-\frac{\tau_1}{\tau_2} + \frac{\tau_1(\tau_1^2 - 1)}{\tau_2^4} + \ldots \\ 1 - \frac{\tau_1^2}{2\tau_2^4 + \ldots }  
	\end{pmatrix} \, , \qquad 
	v_2 = \begin{pmatrix} 1 - \frac{\tau_1^2}{2\tau_2^4} + \ldots \\
	\frac{\tau_1}{\tau_2} - \frac{\tau_1(\tau_1^2 - 1)}{\tau_2^4} + \ldots    
	\end{pmatrix} \, .
}{eq:dec3}
Therefore Eq. \eqref{eq;dec2} implies
\eq{
\begin{split}
v_1^F F + v_1^H H &= \cO(1/\tau_2^2) F + \cO(1) H = \cO(1/\sqrt{\tau_2})\, , \\
v_2^F F + v_2^H H &= \cO(1) F + \cO(1/\tau_2^2) H = \cO(\sqrt{\tau_2})\, .
\end{split}
}{eq:dec4}
While the second equation allows for diverging $F$ and $H$, the first equation implies $H=0$ for $\tau_ 2$ large enough, 
thus ruling out infinite sequences of vacua in this limit.

\subsection{Series approaching a conifold point}

Another commonly occurring kind of singularity in Calabi--Yau manifolds
are conifold singularities. Let us address the
question whether there can be infinite sequences of ISD vacua
accumulating to conifold points\footnote{Close to a conifold point warping effects are large --- see \cite{Douglas:2007tu, Douglas:2008jx}
for the functional form of the corrections to the K\"ahler potential --- and a complete analysis should take also this into account.}.
Consider again our one-parameter models. Around the conifold point
$z=1$ the periods have expansions
\eq{
\begin{pmatrix} \Pi_3 \\ \Pi_2 \\ \Pi_1 \\ \Pi_0  \end{pmatrix} =
\begin{pmatrix} \xi \\ c_0 + c_1\, \xi + \ldots \\ b_0 + b_1\, \xi +
\ldots \\ \frac{\xi}{2\pi i}\log(-i\xi) + a_0 + a_1 \, \xi + \ldots
\end{pmatrix}\, ,
}{eq:con1}
where $\xi \sim (z-1)$. Computing the corresponding metric $\cG_\xi$
produces a matrix with the leading behaviour
\eq{
\cG_\xi \sim \begin{pmatrix} -\frac{2\pi}{\ln |\xi|} + \frac{c_{11}}{\ln^2 |\xi|} &
\frac{b_{12}}{\ln |\xi|} + \frac{c_{12}}{\ln^2 |\xi|} &
\frac{b_{13}}{\ln |\xi|} + \frac{c_{13}}{\ln^2 |\xi|} &
\frac{b_{14}}{\ln |\xi|} + \frac{c_{14}}{\ln^2 |\xi|} \\
\cdot & a_{22} + \frac{b_{22}}{\ln |\xi|} &
a_{23} + \frac{b_{23}}{\ln |\xi|} &
a_{24} + \frac{b_{24}}{\ln |\xi|} \\
\cdot & \cdot &
a_{33} + \frac{b_{3}}{\ln |\xi|} &
a_{34} + \frac{b_{34}}{\ln |\xi|}\\
\cdot &\cdot & \cdot  &
d_{44} \ln|\xi| + a_{44}
       \end{pmatrix} \, ,
}{}
where $a_{ij}$, $b_{ij}$, $c_{ij}$ and $d_{44}$ are constants that are
determined in terms of the expansion coefficients of the periods.
The eigenvectors and eigenvalues of this matrix have the following expansions
\begin{align}\label{eq:eigencon}
&\lambda_1 = -\frac{2\pi}{\ln |\xi|} + \cO(\ln^{-2}|\xi|)\, , 
&w_1^T = & \left[1,\, \cO(\ln^{-1}|\xi|),\, \cO(\ln^{-1}|\xi|) ,\,
\cO(\ln^{-1}|\xi|) \right]\, , \nonumber\\
&\lambda_2 = \ell_A + \cO(\ln^{-1}|\xi|)\, ,
& w_2^T =& \left[\cO(\ln^{-1}|\xi|) ,\, u^A_1 ,\, u^A_2 ,\,
\cO(\ln^{-1}|\xi|) \right]\, , \nonumber\\
&\lambda_3 = \ell_B + \cO(\ln^{-1}|\xi|)\, ,
&w_3^T =& \left[\cO(\ln^{-1}|\xi|) ,\, u^B_1 ,\, u^B_2 ,\,
\cO(\ln^{-1}|\xi|) \right]\, , \nonumber\\
&\lambda_4 = d_{44} \ln |\xi| + \cO(1)\, ,
&w_4^T =& \left[ \cO(\ln^{-2}|\xi|) , \, \cO(\ln^{-1}|\xi|) , \,
\cO(\ln^{-1}|\xi|) ,\,  1 \right] \, .
\end{align}
Here $(u^{A/B})$ and $\ell_{A/B}$ are the eigenvectors and eigenvalues
of the two-by-two matrix
\eq{
\begin{pmatrix} a_{22} & a_{23} \\
a_{23} & a_{33}
       \end{pmatrix}\, ,
}{}
respectively. With these expansions it is straightforward to prove,
in complete parallel to the LCS case, that the flux vectors
$F = (F^0, F^1, F^2, F^3)$ and $H$ must satisfy
\eq{
F^3 = H^3 = 0, \qquad F^{1,2}, H^{1,2} = \cO(1) \qquad {\rm and}
\qquad F^0, H^0 = \cO(\ln^{1/2}|\xi|)\, ,
}{eq:rest}
as $\xi\to 0$ to be able to support an ISD vacuum. (Note that $F^3$ and $H^3$ are fluxes piercing the shrinking cycle.) At this stage, letting $F^0$ and $H^0$ go to infinity produces no contradiction. 
Thus, the simple argument that
disproved infinite sequences in the LCS case is not
sufficient for doing the same for the conifold limit. However,
computing $D_\xi W$ explicitly shows that no infinite series is
possible. To see this we note first that $\tau$ is given by
\eq{
\tau = \frac{F\cdot \Pi^\dagger}{H \cdot \Pi^\dagger} = \frac{F^1
\bar{c}_0 + F^2 \bar{b}_0}{H^1 \bar{c}_0 + H^2 \bar{b}_0} + \cO(\xi)\,.
}{}
To compute $D_\xi W$ we first record the expressions for $K_\xi$ and
$D_\xi \Pi$:
\eq{
K_\xi = \frac{\bar{a}_0 - c_1 \bar{b}_0 + b_1 \bar{c}_0}{2i\,
\im(c_0\bar{b}_0)} + \cO(\xi \log \xi)\, , \qquad
D_\xi \Pi =
\begin{pmatrix} 1 + \cO(\xi) \\ A_1 + \cO(\xi\log \xi)\\A_2 + \cO(\xi\log \xi)
\\ \frac{1}{2\pi i}\log(-i\xi) + \cO(1)
\end{pmatrix}\, ,
}{eq:trouble_ahead}
with
\eq{
A_1 = c_1 + c_0 \, \frac{\bar{a}_0 - c_1 \bar{b}_0 + b_1
\bar{c}_0}{2i\, \im(c_0\bar{b}_0)} \qquad
A_2 = b_1 + b_0 \, \frac{\bar{a}_0 - c_1 \bar{b}_0 + b_1
\bar{c}_0}{2i\, \im(c_0\bar{b}_0)} \, .
}{eq:trouble_behind}
This yields
\eq{
\begin{split}
D_\xi W  &= F^0 - \frac{F^1 \bar{c}_0 +
F^2 \bar{b}_0}{H^1 \bar{c}_0 + H^2 \bar{b}_0} H^0 +
\frac{(F^1 H^2 - F^2 H^1)\bar{a}_0}{\bar{c}_0 H^1 + \bar{b}_0 H^2}
+ \cO(\xi\ln|\xi|)\\
&= F^0 - \tau H^0 + \cO(1)
\, .
\end{split}
}{}
We see from this expressions that in order to have $D_\xi W = 0$ as $F^0, H^0
\to \infty$, $\tau$ must approach the real ratio $F^0/H^0$. This means that
the
imaginary part of $\tau$ goes to zero, which is S-dual to a decoupling
limit. Therefore, as in the LCS case, there are no infinite
sequences of vacua with finite (and nonzero) string coupling.

\subsection{The two-paramter model $\mathcal{M}_{(86,2)}$}
Until now we have studied D-limits in the complex structure and axio-dilaton moduli spaces of  a family of one-parameter Calabi--Yau manifolds.
In this section, as a first step to a more general result, we extend the previous result to a specific two-parameter 
model. Again we find that there is no infinite sequence of supersymmetric vacua 
approaching the LCS point. 

Consider the two-parameter model $\mathcal{M}_{(86,2)}$. Its periods can be 
expanded around the LCS point \cite{Chialva:2007sv}:

\begin{gather}
\begin{pmatrix}
	\Pi_5 \\ \Pi_4 \\ \Pi_3 \\ \Pi_2 \\ \Pi_1 \\ \Pi_0 \\
\end{pmatrix}
\sim
\begin{pmatrix}
	\delta_3 - \frac{25}{12}\,y - x - \frac{1}{6}\left( 5 y^3 + 12 y^2 x\right) \\
	-\frac{25}{12} + \frac{1}{2} y + \frac{5}{2} y^2 + 4 x y \\
	-1 + 2 y^2\\
	y\\
	x\\
	1\\
\end{pmatrix}
\, .
\end{gather}
Here $\delta_3=\frac{21 i \zeta(3)}{\pi^3}$ is a constant, whereas $x$ and $y$ 
are the two complex structure moduli. 
Approaching the LCS point corresponds to sending $\text{Im}\, x\rightarrow 
\infty$ and $\text{Im}\, y\rightarrow \infty$, where the limits can be taken 
independently. 
The K\"ahler potential takes the form
\begin{equation}
	e^{-K}=16 \, x_2\, y_2^2+\frac{20}{3} y_2^3 +\frac{42  \,\zeta(3)}{\pi^3} + \ldots,
\end{equation}
where $x=x_1+i x_2$ and $y=y_1+i y_2$ with $x_i,y_i\in\mathbb{R}$. 
 
Consider the case in which the limits for the two variables are taken at the 
same time, i.e. $x_2=y_2=z \rightarrow \infty$. 
This choice significantly simplifies the computation of the metric $\mathcal{G}_z$, which results in
{\footnotesize
\begin{gather}\label{eqn:Gz_twoparameter}
\mathcal{G}_z	= 
\begin{pmatrix}
	a_{11} \,z^3 + \cO(z) & a_{12} \,z +\cO(\frac{1}{z}) & a_{13}\, z+\cO(\frac{1}{z}) & a_{14}\, \frac{1}{z}+\cO(\frac{1}{z^2}) &  a_{15}\, \frac{1}{z} +\cO(\frac{1}{z^2})&  a_{16}\, \frac{1}{z^3}+\cO(\frac{1}{z^4}) \\
	\cdot & a_{22}\, z +\cO(\frac{1}{z}) & a_{23}\, z +\cO(\frac{1}{z}) & a_{24}\, \frac{1}{z}+\cO(\frac{1}{z^3}) &  a_{25} \,\frac{1}{z}+\cO(\frac{1}{z^3}) &  a_{26}\, \frac{1}{z^3}+\cO(\frac{1}{z^4}) \\	 		 
	\cdot & \cdot & a_{33}\, z +\cO(\frac{1}{z})& a_{34}\, \frac{1}{z} +\cO(\frac{1}{z^3})&  a_{35}\, \frac{1}{z} +\cO(\frac{1}{z^3})&  a_{36}\, \frac{1}{z^3} +\cO(\frac{1}{z^4})\\
	\cdot & \cdot & \cdot & a_{44}\, \frac{1}{z}+\cO(\frac{1}{z^3}) &  a_{45}\, \frac{1}{z}+\cO(\frac{1}{z^3}) &  a_{46}\, \frac{1}{z^3}+\cO(\frac{1}{z^6}) \\
	\cdot & \cdot & \cdot & \cdot &  a_{55}\, \frac{1}{z} +\cO(\frac{1}{z^3})&  a_{56}\, \frac{1}{z^3}+\cO(\frac{1}{z^6}) \\
	\cdot & \cdot & \cdot & \cdot & \cdot &  a_{66}\, \frac{1}{z^3}+\cO(\frac{1}{z^6})
	\end{pmatrix}
	\, .
\end{gather}}
The constants $a_{ij}$ are known, and we collect them in Appendix \ref{App2para}.
The eigenvectors and eigenvalues of this metric expanded in $z$ are given by
\begin{align}\label{eqn:eigen2}
&\lambda_1 = a_{11} z^3 +\cO(z)\,, & w_1^T=&\left[1,\cO(z^{-2}), \cO(z^{-2}),\cO(z^{-4}),\cO(z^{-4}),\cO(z^{-6})\right]\nonumber \\
&\lambda_2 = \ell_{2} z +\cO(z^{-1})\,, & w_2^T=&\left[\cO(z^{-2}), w_2^2+\cO(z^{-2}), w_2^3+\cO(z^{-2}),\cO(z^{-2}),\cO(z^{-2}),\cO(z^{-4})\right]\nonumber \\
&\lambda_3 = \ell_{3} z +\cO(z^{-1})\,, & w_3^T=&\left[\cO(z^{-2}), w_3^2 + \cO(z^{-2}),w_3^3+\cO(z^{-2}),\cO(z^{-2}),\cO(z^{-2}),\cO(z^{-4})\right]\nonumber \\
&\lambda_4 = \frac{\ell_{4}}{z} +\cO(z^{-3})\,, & w_4^T =& \left[\cO(z^{-4}),  \cO(z^{-2}),  \cO(z^{-2}),  w_4^4 + \cO(z^{-2}), w_4^5 + \cO(z^{-2}),  \cO(z^{-2})\right] \nonumber\\
&\lambda_5 = \frac{\ell_{5}}{z} +\cO(z^{-3})\,, &w_5^T =& \left[\cO(z^{-4}),  \cO(z^{-2}),  \cO(z^{-2}), w_5^4+ \cO(z^{-2}), w_5^5+ \cO(z^{-2}), \cO(z^{-2})\right] \nonumber \\
&\lambda_6 = \frac{a_{66}}{z^3} +\cO(z^{-6})\,, & w_6^T=&\left[\cO(z^{-6}), \cO(z^{-4}),\cO(z^{-4}),\cO(z^{-2}),\cO(z^{-2}),1\right]
\, .
\end{align}
The $\ell_i$ and $w_i^j$ are the eigenvalues and eigenvectors of
appropriate two-by-two matrices. Consider the flux-vector $N_n=(N^0_n,\dots,N^5_n)$ and the following limit:
\begin{equation}
\lim_{n\rightarrow\infty}  \, z_n =\infty \qquad \lim_{n\rightarrow\infty} N_n\cdot \cG_{z_n} \cdot N^T_n\equiv \lim_{z\rightarrow\infty}
\sum_{i=1}^6 \vert  N \cdot  w_i(z)  \vert ^2 \lambda_i(z) \neq\infty \, .
\end{equation}
In order not to clutter notation, we will suppress the index $n$ in the following.

Without loss of generality assume $N^0=1$.
The first eigenvalue grows as $\lambda_1\sim z^3$. Therefore we must have $| N  \cdot w_1|^2 \sim\cO(z^{-3})$. Hence
\begin{equation}
 N \cdot w_1 \sim \cO(z^{-3/2}) = 1 + N^1 \cO(z^{-2})+N^2\cO(z^{-2})+N^3\cO(z^{-4})+N^4\cO(z^{-4})+N^5\cO(z^{-6}) \, .
\end{equation}
This can happen only if at least one of the $N^i$ diverges. It is also clear that this must happen in order for $N$ to approach one
of the zero eigenvectors of $\cG_z$. What is needed is 
\begin{align}
&N^1 = P z^2 + o(z^2)\, ,  &N^2 &= Q z^2+o(z^2)\, , \nonumber \\
&N^3 = R\, z^4 +o(z^4)\, , &N^4 &= S z^4 +o(z^4)\, , \nonumber \\ 
&N^5 = T\, z^6 +o(z^6) \, .
\end{align}
Recall that $o(z^2)$ stay for terms that grow slower than $z^2$ and $P,Q,R,S,T$ are appropriate constants.

Consider
\begin{equation}
\cO(z^{3/2}) =  N \cdot w_6 = T z^6 + S \cO(z^2) + R \cO(z^2) +  \ldots
\end{equation}
This immediately proves that $T$ must be zero. We set it to zero in the following. Furthermore, consider
\begin{equation*}
\cO(z^{1/2}) =  N \cdot w_4  = (R w_4^4 + S w_4^5)z^4 + \ldots
\end{equation*}
\begin{equation}
\cO(z^{1/2}) =  N \cdot w_5  = (R w_5^4 + S w_5^5)z^4 + \ldots
\end{equation}
This proves that $(R w_4^4 + S w_4^5) = (R w_5^4 + S w_5^5) = 0$, i.e., 
\begin{align}
\begin{pmatrix} w_4^4 & w_4^5 \\ w_5^4 & w_5^5  \end{pmatrix} \begin{pmatrix} R\\ S  \end{pmatrix} = 0.
\end{align}
Since the matrix is orthogonal it follows that $R = S = 0$. 

We continue the analysis along the line of the LCS case of the one-parameter models. In the end we obtain following conditions on the flux-vectors $F=(F^0,\dots,F^5)$ and $H=(H^0,\dots, H^5)$:
\begin{equation}
F^0=F^1=F^2=0 \qquad \text{and} \qquad H^0=H^1=H^2=0\;.
\end{equation}
This result means that there is no ISD vacuum approaching the LCS. The only exception is again the singular vacuum located 
exactly at the LCS point.

\section{D-limits and infinite flux series for F-theory on $K3\times K3$}
\label{sec:dlimK3}

The simplest non-trivial flux compactifications apart from toroidal orbifolds are compactifications of type 
IIB string theory on the orientifold $K3\times T^2/\mathbb{Z}_2$. These models contain four orientifold planes and 
16 D7-branes which are points in $T^2/\mathbb{Z}_2$ and fill out the entire $K3$ as well as the four non-compact directions.

Alternatively, these compactifications can be described as F-theory on $K3\times K3$. This description not only allows 
for an elegant treatment of IIB flux compactifications on the orientifold $K3\times T^2/\mathbb{Z}_2$, but also naturally 
includes two-form fluxes on the D7-branes. As shown by Aspinwall and Kallosh \cite{Aspinwall:2005ad}, the number of 
supersymmetric vacua of such compactifications is finite, i.e. there can be no infinite flux series in these models. 

In this section, we discuss this result from the perspective of D-limits. We are able to find D-limits as well as associate 
infinite flux sequences on $K3$, so that the result of \cite{Aspinwall:2005ad} implies that all but finitely many of the 
corresponding solutions are actually equivalent by automorphism of the lattice $H^2(K3,\mathbb{Z})$. 
We demonstrate this in a simple example.

Compactifications of type IIB string theory on $K3\times T^2/\mathbb{Z}_2$ with $G_{3}$ flux have also been considered 
in \cite{Tripathy:2002qw}. They show how to find an infinite sequence of fluxes which solves all of the supersymmetry
conditions except for primitivity. In general, imaginary self-duality (ISD) does not imply supersymmetry. In the 
present case, however, one can show that the complex structure of $K3$ may always be chosen such that (for fixed metric) 
the supersymmetry constraints are satisfied for any ISD solution. Hence their sequence also breaks imaginary 
self-duality. Therefore, we can not treat the flux series of \cite{Tripathy:2002qw} as a D-limit in the sense introduced.

\subsection{F-theory with $G_{(4)}$ flux on $K3\times K3$}

In compactifications of F-theory on the fourfold $K3\times K3$ \footnote{One of the $K3$s has to be elliptically fibered for F-theory to make 
sense. We assign tildes to quantities associated with the elliptic $K3$.}, one can switch on four-form fluxes $G_{(4)}$ which are 
integrally quantized. They can be written as
\begin{equation}
 G_{(4)}=G^{\mu\nu}\eta_\mu\wedge\tilde{\eta}_\nu \, .
\end{equation}
Here $\eta_\mu$ and $\tilde{\eta}_\nu$ are integral two-forms on the two $K3$s. We will think of the
matrix $G^{\mu\nu}$ as the components of a vector in $H^2(K3,\Z)\otimes H^2(K3,\Z)$ and simply write
$G$ in the following.

The scalar potential induced by the fluxes can stabilize both complex structure as well as
K\"ahler moduli of $K3\times K3$ (except for the volumes of the two $K3$s). The vacua of these
models were analysed in \cite{Aspinwall:2005ad, Braun:2008pz}. See also \cite{Tripathy:2002qw, Andrianopoli:2003jf} for 
an analysis from the type IIB perspective.

It can be shown that the scalar potential is positive definite and can be written as \cite{Braun:2008pz} 
\begin{equation}
 V=\frac{1}{2}\int_{K3\times K3}G_{(4)}\wedge(\ast G_{(4)} - G_{(4)}) \, .
\end{equation}
As $G_{(4)}$ is forced to be self-dual by the equations of motion, their solutions correspond to Minkowski minima 
of the effective potential. 

For $K3\times K3$, the tadpole condition \eqref{Mtadpole} reads
\begin{equation}\label{k3tadpole}
 \frac{1}{2}\int_{Y_4}G_{(4)}\wedge G_{(4)} +N_{D3} = 24 \, ,
\end{equation}
which can be rewritten as
\begin{equation} 
24=N_{D3} + \frac{1}{2}\int_{K3\times K3}G_{(4)} \wedge G_{(4)}=N_{D3} +\frac{1}{2}\int_{K3\times K3}G_{(4)} \wedge\ast G_{(4)} \, ,
\end{equation}
which is manifestly positive. In order to discuss D-limits, we consider the metric $\mathcal{G}$, which is defined by
\begin{equation}
 G \cdot \mathcal{G}\cdot G^T \equiv \frac{1}{2}\int_{K3\times K3}G_{(4)} \wedge\ast G_{(4)} \, .
\end{equation}
As we consider a fourfold which is a product of two spaces, we can decompose
\begin{equation}
 \mathcal{G}= \frac{1}{2}\mathcal{G}_{\Sigma}\otimes  \mathcal{G}_{\tilde{\Sigma}} \, .
\end{equation} 
Given an integral two-form $G_{(2)}=\sum_\mu g_\mu\eta^\mu$, $\mathcal{G}_{\Sigma}$ is defined by 
\begin{equation}\label{defGsigma}
 \int_{K3}G_{(2)} \wedge\ast G_{(2)}=g\cdot  \mathcal{G}_{\Sigma}\cdot g^T \, .
\end{equation}
As before, a D-limit is defined to be a limit in moduli space in which $\mathcal{G}$, i.e. $\mathcal{G}_{\Sigma}$
or $\mathcal{G}_{\tilde{\Sigma}}$, degenerates.

\subsection{The $K3$ surface}\label{sect:k3}

In order to discuss the properties of $\mathcal{G}_{\Sigma}$ and find which D-limits we can
have for F-theory compactification on $K3\times K3$ we collect a few crucial properties 
about $K3$ surfaces in this section. For a more thorough treatment, see e.g. \cite{Aspinwall:1996mn,peters}.

In two complex dimensions there is just one non-trivial compact Calabi--Yau manifold: $K3$. 
The metric deriving from the natural inner product on the $22$-dimensional space $H^2(K3)$:
\begin{equation}\label{prodk3}
M_{\mu\nu}=\int_{K3}\eta_\mu\wedge\eta_\nu,
\end{equation}
has signature $(3,19)$. 
The vector space $H^2(K3)$ contains the lattice $H^2(K3,\mathbb{Z})$, the elements of which are Poincar\'e dual
to curves in $K3$. This lattice can be written as 
\begin{equation}\label{k3lattice}
 H^2(K3,\mathbb{Z})=-E_8^{\oplus 2}\oplus U^{\oplus 3} \, ,
\end{equation}
where $E_8$ denotes the root lattice of $E_8$ and $U$ is the hyperbolic lattice. Embedded in
a vector space with orthonormal basis $E_I$, the root lattice of $E_8$ is given by vectors 
\begin{equation}
\sum_I q_I E_I \, ,
\end{equation}
where the $q_{I}$ have to be \textit{all} integer or \textit{all} half-integer and
fulfill the relations $\sum_{I=1,...,8}q_{I} \in 2\mathbb{Z}$. The lattice $U^{\oplus 3}$ is spanned
by integral multiples of $e_i$, $e^j$ which have the intersections
\begin{equation}
 e_i\cdot e^j=\delta^j_i \qquad e_i\cdot e_j=0\qquad e^i\cdot e^j=0 \, .
\end{equation}
The inner product between integral two-forms has a geometric interpretation as the intersection
of the dual curves. As the $K3$ surface has a trivial canonical bundle, the self-intersection number 
of a curve, i.e. the intersection between two homologous curves, translates to its genus by using 
the adjunction formula. Denoting the curve dual to the integral two-form $\eta_C$ by $C$ one obtains
\begin{equation}
\int_{K3}\eta_C\wedge \eta_C=C\cap C=-\chi(C)=-2+2g(C) \, . 
\end{equation}
The geometric moduli space of $K3$ is the set of all oriented positive-norm three-planes $\Sigma$ in 
$H^2(K3)$ modulo automorphisms of the lattice $H^2(K3,\mathbb{Z})$ in $O^+(3,19)$ \cite{Aspinwall:1996mn, autK3}. 
The group $O^+(3,19)$ is the component of the orthogonal group which leaves the orientation
of $\Sigma$ invariant.
 
We span $\Sigma$ using three orthonormal vectors $\omega_i$:
\begin{equation}\label{condomegai}
 \omega_i\cdot\omega_j =\delta_{ij} \, .
\end{equation}
Note that this description leaves an $SO(3)$ symmetry, rotating the $\omega_i$ into one
another. We can construct the K\"ahler form $J$ and the holomorphic two-form $\Omega$
of $K3$ using the vectors $\omega_i$:
\begin{equation}\label{OJofw}
 \Omega=\omega_1+i\omega_2 \qquad J=\sqrt{V} \omega_3 \, ,
\end{equation}
where we have denoted the volume of $K3$ by $V$. It is important to note that a choice of $\Sigma$ 
determines the metric of $K3$ (up to the overall volume), but does not completely
determine the complex structure. We still may rotate the $\omega_i$ inside $\Sigma$ or 
equivalently change the definition in \eqref{OJofw}. For a fixed complex structure, the lattice of 
integral cycles of $K3$ which are orthogonal to $\Omega$ is the Picard lattice.

Any two-form $H \in H^2(K3)$ can be decomposed into a piece parallel and
a piece perpendicular to $\Sigma$:
\begin{equation}
 H=H_{\parallel}+H_{\perp} \, .
\end{equation}
The action of the Hodge-$\ast$ operation on $K3$ then takes the simple form
\begin{equation}
 \ast H=\ast H_{\parallel}+\ast H_{\perp}=H_{\parallel}-H_{\perp} \, .
\end{equation}

The $K3$ moduli space naturally includes loci over which the $K3$ surface develops ADE
singularities. Whenever there are elements $\gamma_i\in H^2(K3,\mathbb{Z})$ with 
$\gamma_i\cdot\gamma_i=-2$ that are orthogonal to $\Sigma$, the dual spheres collapse to 
produce an ADE singularity. Loci where this occurs are at a finite distance in moduli 
space from any generic smooth $K3$.

\subsection{D-limits and $\mathcal{G}_\Sigma$}

Let us now see if $\mathcal{G}_\Sigma$ can degenerate so that we find a D-limit.
As before, we denote the vector of coefficients that is obtained when an integral two-form $G_{(2)}$ is expanded in some 
basis $\eta_{\mu}$ of $H^2(K3)$ by $g$. In order to facilitate the discussion of flux quantization we
choose this basis to be integral, i.e. the vectors $\eta_{\mu}$ are elements of the lattice $H^2(K3,\mathbb{Z})$.

Using this basis,
\begin{align}
\omega_i=\omega_i^{\mu}\eta_{\mu} \, ,\qquad G_{(2)}=g^{\mu}\eta_{\mu} \, .
\end{align}
We can decompose 
\begin{align}\label{Gdecomp}
 G_{(2)}=G_{(2)\parallel}+G_{(2)\perp}=\sum_i \int_{K3}(\omega_i\wedge G_{(2)})\, \omega_i + G_{(2)}-\sum_i \int_{K3}(\omega_i\wedge G_{(2)})\omega_i \, .
\end{align}
Hence
\begin{align}
 \ast G_{(2)}&=G_{(2)\parallel}-G_{(2)\perp}=\left( 2 M_{\mu\nu}\sum_i \omega_i^{\rho}\omega_i^{\mu}-\delta_{\nu}^{\rho} \right) g^{\nu}\eta_{\rho}\, .
\end{align}
Defining the projector
\begin{equation}
 \Gamma^{\rho}_\nu\equiv \sum_i \omega_i^{\rho}\omega_i^{\mu}M_{\mu\nu}\qquad \Gamma^2=\Gamma \, ,
\end{equation}
which projects any form onto its components parallel to $\Sigma$, we obtain
\begin{equation}\label{Gsigma}
 \mathcal{G}_{\Sigma}=M\left(2\Gamma-1\right) \, .
\end{equation}
As the inner product \eqref{defGsigma} is positive definite for any smooth $K3$ surface, it follows
that the metric in \eqref{Gsigma} has the same property. It can only degenerate in a limit in moduli
space in which the $K3$ surface becomes singular. Let us first consider the aforementioned ADE singularities.
They occur when we rotate the three-plane $\Sigma$ such that it becomes orthogonal to specific lattice 
vectors of $H^2(K3,\mathbb{Z})$ with respect to the metric \eqref{prodk3}. The expression we have derived
for $\mathcal{G}_\Sigma$, however, does not at all depend on the location of $\Sigma$ relative to 
the lattice $H^2(K3,\mathbb{Z})$. Hence the metric $\mathcal{G}_\Sigma$ can not degenerate when we
approach a locus in moduli space for which the $K3$ surface has an ADE singularity. Note also that these 
singularities occur at finite distance in moduli space, i.e. the naturally lie \emph{inside} the moduli space.
Another kind of singularity occurs when we rotate $\Sigma$ towards a light-like direction in $H^2(K3)$. 
In the following, we shall investigate such a limit and show that it indeed gives rise to a degeneration
of $\mathcal{G}_\Sigma$.

A well-known example of such a limit is the F-theory limit of a compactification of
M-theory on an elliptically fibered $K3$. In this limit, the volume of the $T^2$ fibre is 
taken to zero, which corresponds to rotating the K\"ahler form towards the light cone
in $H^2(K3)$ \cite{Braun:2009wh}. Just as in the case of the large complex structure 
limit, this limit is dual to a decompactification limit which takes place on the type 
IIB/F-theory side. Furthermore, it can be shown that this limit is at infinite distance
in moduli space \cite{Braun:2008pz}.

\subsubsection*{An example}

To show that rotating $\Sigma$ towards the light cone constitutes a D-limit, we consider a simple example. 
For ease of exposition, we keep the three-plane $\Sigma$ in a four-dimensional subspace spanned by
\begin{equation}
d^1=e_1+e^1 \, , \qquad d^2=e_2+e^2 \, ,\qquad d^3=e_3+e^3 \, ,\qquad d^4=e_1-e^1 \, .
\end{equation}
Note that the intersection form is diagonal in terms of the $d^i$: $M=\text{diag}(2,2,2,-2)$.
In this basis, we choose $\Sigma$ to be spanned by the orthonormal vectors
\begin{equation} \label{K3Gseries}
 \omega_1(n)=\frac{1}{\sqrt{2}}(1,0,n,n),  \qquad
 \omega_2(n)=\frac{1}{\sqrt{2}}(0,1,0,0), \qquad
 \omega_3(n)=\frac{1}{\sqrt{2+2n^2}}(n,0,-1,0) \, .
\end{equation}
The matrix $\mathcal{G}_{\Sigma}=M(2\Gamma-1)$ is 
\begin{equation}
 \mathcal{G}_{\Sigma}=2\begin{pmatrix}
                       \frac{1+3n^2}{1+n^2} & 0\qquad & \frac{2n^2}{1+n^2} & -2n  \\
		      0 		    & 1\qquad &    0               &    0 \\
		      \frac{2n^2}{1+n^2} & 0\qquad &-1+2(n^2+\frac{1}{1+n^2}) & -2n^2  \\
		      -2n               & 0\qquad & -2n^2		& 1+2n^2   
                      \end{pmatrix}\, .
\end{equation}
Its eigenvalues are given by
\begin{equation}
\lambda_1= \lambda_2=2\, , \qquad \lambda_3=2+4n^2-4\sqrt{n^2+n^4},  \qquad \lambda_4=2+4n^2+4\sqrt{n^2+n^4}\, .
\end{equation}
In the limit $n\rightarrow\infty$ we may approximate
\begin{equation}
 \sqrt{n^2+n^4}= n^2+\frac{1}{2}-\frac{1}{8n^2}+\mathcal{O}(n^{-4}) \, .
\end{equation}
In this limit we hence find that $\lambda_3$ goes to zero and $\lambda_4$ goes to infinity:
\begin{equation}
 \lambda_3\sim \frac{1}{2n^2} \qquad \lambda_4 \sim 8n^2 \, .
\end{equation}
The eigenvector $v_3$ associated with $\lambda_3$ is
\begin{equation}
v_3= \left(
\begin{array}{c}
 \frac{-n^3+n \left(-1+\sqrt{n^2+n^4}\right)}{\left(1+n^2\right) \left(n^2-\sqrt{n^2+n^4}\right)} \\
 0 \\
 -\frac{n^2 \left(1+n^2-\sqrt{n^2+n^4}\right)}{\left(1+n^2\right) \left(n^2-\sqrt{n^2+n^4}\right)} \\
 1
\end{array}
\right) \overrightarrow{n\rightarrow\infty} \left( \begin{array}{c}n^{-1}\\0\\1\\1 \end{array} \right) \, .
\end{equation}
Hence we find a D-limit in which the metric degenerates in the direction of $\omega_1$.
Note that this is precisely the direction of $\Sigma$ which we are rotating towards the light-cone.

Let us now use this example to construct a flux series. The flux vector
\begin{equation}
g= (1,0,n,n,0,0,...)\sim n v_3 \, ,
\end{equation}
is properly quantized for any integer $n$. For large values of $n$ we have that   
\begin{equation}
g \cdot \mathcal{G}_{\Sigma}\cdot g^T \sim n^2 \frac{1}{n^2} = \text{const} \, .
\end{equation}
Hence the eigenvalue of $v_3$ goes to zero fast enough to allow for an infinite 
sequence of integral flux vectors for which $g\cdot\mathcal{G}_\Sigma\cdot g^T$ approaches a constant
in the D-limit.

\subsection{Infinite series and automorphisms of $H^2(K3,\mathbb{Z})$}

To put the example of the last section to work we set
\begin{align}
 \Omega(n)=\omega_1(n)+i\omega_2   \\
 \tilde{\Omega}=\tilde{\omega}_1+i \tilde{\omega}_2   \, ,
\end{align}
with $\omega_i(n)$ given by \eqref{K3Gseries}. A properly quantized flux series that obeys the supersymmetry conditions 
(and equations of motion) is given by \cite{Aspinwall:2005ad}
\begin{equation}
 G_{(4)}(n)=\sqrt{2}\left( \omega_1(n)\wedge\tilde{\omega}_1+\omega_2\wedge\tilde{\omega}_2 \right) \, .
\end{equation}
The flux-induced D3 tadpole is
\begin{equation}
  \frac{1}{2}\int_{K3\times K3}G_{(4)}(n) \wedge G_{(4)}(n)=2 \, ,
\end{equation}
for any $n$. For the K\"ahler form $J$ we can choose any positive norm two-form in $H^2(K3)$ which is
orthogonal to $\Omega$. Setting $J=\omega_3(n)$ demonstrates that such a $J$ can always be found.

In \cite{Aspinwall:2005ad}, it was shown that there can only be a finite number of supersymmetric flux vacua in 
compactifications on $K3\times K3$. In order to make contact with our results, we review their main results.
As supersymmetry demands that the flux is of type $(2,2)$ and primitive, one can write
\begin{equation}\label{fluxaspkall}
 G=\text{Re}\left(c\,\Omega\wedge\bar{\tilde{\Omega}}\right)+\sum_{\alpha}\psi_{\alpha}\wedge\tilde{\psi_{\alpha}} \, ,
\end{equation}
where $c$ is a parameter that has to be chosen appropriately for flux quantization and $\psi_{\alpha}, \tilde{\psi_{\alpha}}$
are integral primitive $(1,1)$ forms on the respective $K3$ surfaces. They show that if only the first term is present, as
is the case for our example, the complex structure moduli of the two $K3$ surfaces, i.e. $\Omega$ and $\tilde{\Omega}$,
are completely fixed. Furthermore, they are fixed such that the Picard lattice of the corresponding $K3$ surfaces is of
maximal rank, i.e. $\Omega$ sits inside a two-dimensional lattice $\Upsilon\subset H^2(K3,\mathbb{Z})$. Such $K3$ surfaces, 
which have been dubbed\footnote{They have also been referred to as `singular' $K3$ surfaces, even though they can
be perfectly smooth manifolds. Hence we follow \cite{Aspinwall:2005ad} in calling them `attractive'.} `attractive', can be classified 
through the lattice $\Upsilon$. It turns out that only a finite number of attractive $K3$ surfaces can satisfy the tadpole
condition \eqref{k3tadpole}. 
When the second term in \eqref{fluxaspkall} is also present, the $K3$ ceases to be attractive. %really ? 
Its contribution to the tadpole is, however, always positive definite. Hence there can be only a finite number of flux choices
that admit supersymmetric flux vacua and satisfy the tadpole condition for F-theory on $K3\times K3$.

Supersymmetry only forces $G_{(4)}$ to be primitive, but does not fix the K\"ahler moduli. Non-perturbative effects, however,
give rise to an effective potential that can fix all K\"ahler moduli. As the effective potential is determined once
fluxes (and hence the complex structure) are given, it follows that there is only a finite number of supersymmetric
stable flux vacua for F-theory on $K3\times K3$. In case both terms in \eqref{fluxaspkall} are non-zero, some
of the instantons that stabilize the K\"ahler moduli can be obstructed, so that not all moduli are stabilized.

The results of \cite{Aspinwall:2005ad} indicate that all but a finite number of the vacua of the series we have constructed 
before must actually be equivalent. Note that for our series, only the first term in \eqref{fluxaspkall} is present. 
Once we specify $G$ in terms of $\Omega$, the K\"ahler form is therefore determined completely. Hence we have to show
that there is an automorphism of $H(K3,\mathbb{Z})$ which identifies all but a finite number of the sub-lattices spanned by 
the $\Omega(n)$. To find this automorphism, we write $\omega_1(n)$ and $\omega_2$ in terms of a basis for the 
lattice $U^{\oplus 3}$:
\begin{align}
\omega_1(n)&=e_1(1+n)+ne_3+e^1(1-n)+ne^3 \, , \\
\omega_2&=e_2+e^2 \, .
\end{align}

Indeed, there is an automorphism of $H^2(K3,\mathbb{Z})$ which identifies \emph{all} of the solution in our series. It is given by
\begin{align}
e_{1} & \mapsto \hat{e}_{1}= (1+n)e_{1}+ne_{3}   & e^{1}\mapsto  \hat{e}^1&=(1-n)e^{1} + ne^{3}\nonumber \\
e_{2} & \mapsto \hat{e}_2=-e_2 &  e^{2} \mapsto \hat{e}^2&=-e^2 \nonumber \\
e_3& \mapsto \hat{e}_{3}=ne_1+(n-1)e_3  & e^{3} \mapsto \hat{e}^3&=ne^1 -(1+n)e^3 \, .
\end{align}
with all other elements unchanged. It maps
\begin{align}
\omega_1(0)=e_1+e^1&\mapsto \hat{e}_1+\hat{e}^1 = \omega_1(n)\, ,\\
\omega_2&\mapsto -\omega_2 \, .
\end{align}
Hence this automorphism identifies the holomorphic two-forms and consequently also the fluxes of our series of $K3$ surfaces.
Furthermore, it gives rise to an orientation preserving\footnote{Note that this is not the case if we leave $e_2$ and $e^2$ invariant.} 
map of $\Sigma$ to itself. Thus it is induced from a diffeomorphism of $K3$ \cite{autK3}, so that all of the solutions in our series should 
be considered equivalent. 

Our example is, of course, very simple in that it only rotates $\Omega$ towards the light cone in the lattice $U^{\oplus 3}$. 
Even though examples of D-limits and infinite flux series employing the $E_8$ lattices can be constructed in a straightforward 
fashion, the corresponding automorphisms are harder to find. Showing that such automorphisms exist for any D-limit would hence 
constitute an alternative proof of the finiteness of the number of supersymmetric flux vacua on $K3\times K3$.
As the self-duality condition on $G_4$ follows from the equations of motion but does not require supersymmetry, one
could then try to prove a similar theorem also for non-supersymmetric vacua.

One can turn this logic around and construct automorphism of $K3$ by studying D-limits. By the result of \cite{Aspinwall:2005ad}, 
only a finite number of solutions in any infinite sequence of supersymmetric vacua can be different. Hence there must be 
corresponding automorphisms in $O^+(3,19)$ which identify all but a finite number of the solutions. It would be interesting to use 
this approach to study the diffeomorphism group of $K3$ surfaces.

As the self-duality condition also holds without supersymmetry, infinite sequences can also only occur in D-limits in this case. With a sufficient 
understanding of the automorphisms of $K3$ it hence seems possible to use the D-limit approach to study the existence of infinite sequences of 
non-supersymmetric solutions. 

\section{The models of Ahlqvist et al.}\label{sec:nummod}

\begin{figure}[tb]
\centering
\subfigure{\includegraphics[height=7cm]{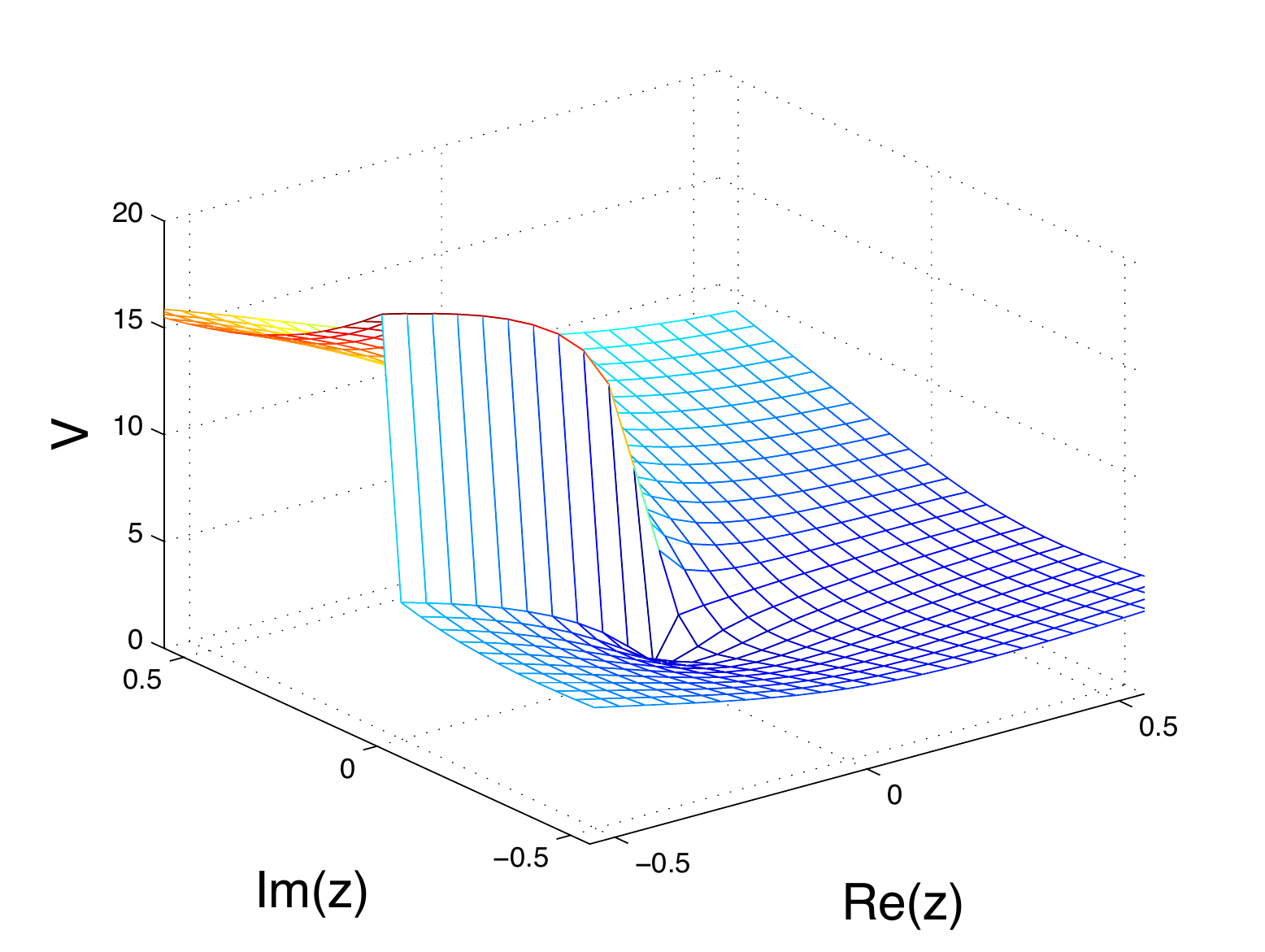}}
\qquad
\subfigure{\includegraphics[height=7cm]{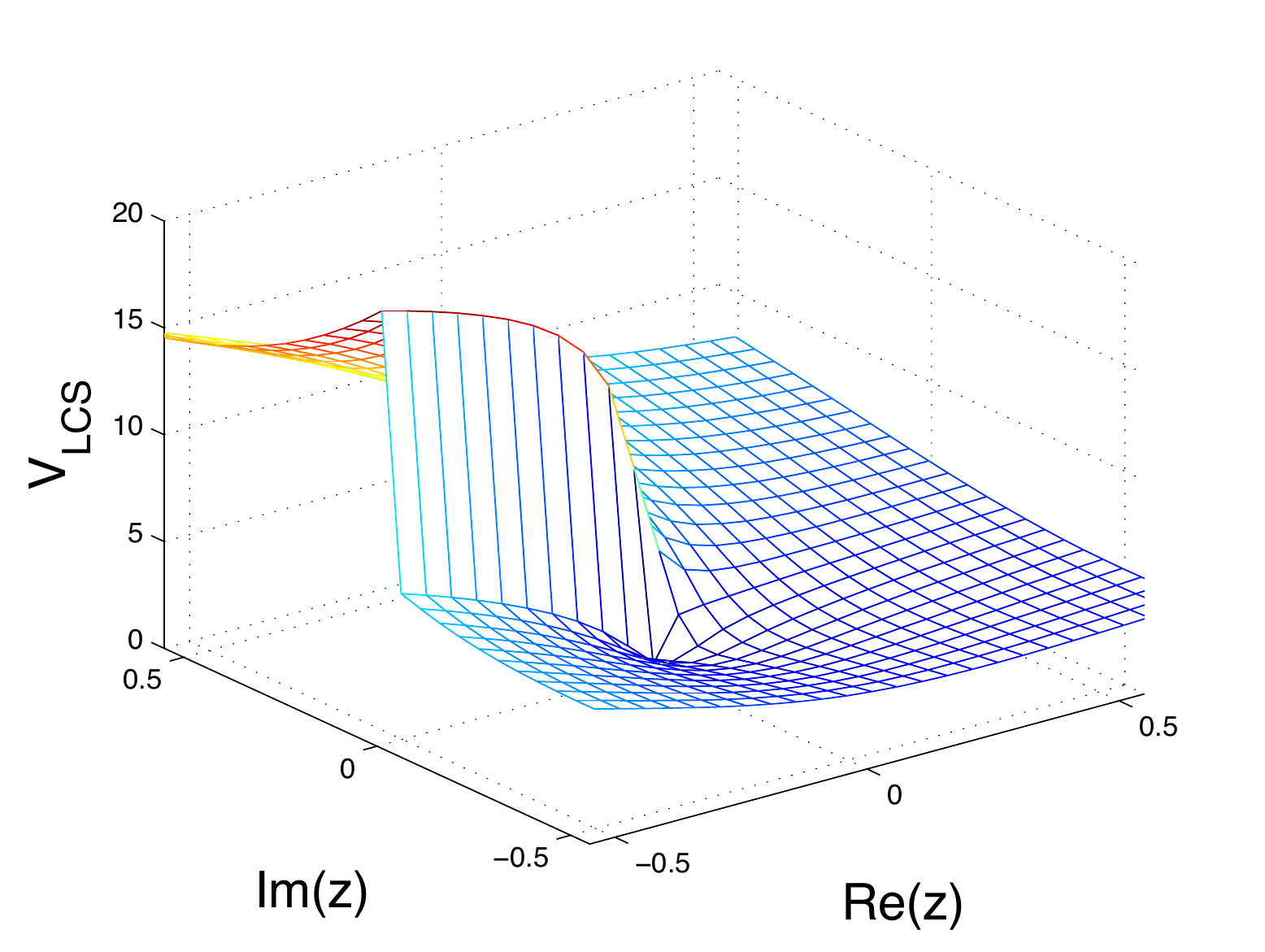}}
\caption{The scalar potential for the complex structure modulus $z$ of the Mirror Quintic with NSNS flux $H=(-2,-4,-33,0)$ and RR flux $F = (3,-18,9,-1)$. The potential has already been minimized with respect to the axio-dilaton $\tau$, so the minimum shown is a minimum for both $z$ and $\tau$. The first panel shows the potential calculated with periods calculated with the full Meijer functions, whereas the second potential is calculated with the LCS expansions of the periods. As can be seen from the figures, the LCS expansions are enough to reproduce the features of this minimum.}
\label{fig:meijer_lcs}
\end{figure}

In this section we take a closer look at  a few examples of sequences of minima that converge to the LCS point, and that were first reported on in \cite{Ahlqvist:2010ki}. These minima have 
vanishing scalar potential and hence fulfill the ISD condition.  A question left open in this reference was whether these series are infinite. Here we use the LCS expansions of the periods 
to show that there are more minima in the series than those reported in \cite{Ahlqvist:2010ki}, but that the minima eventually break the ISD condition and the series terminate in agreement 
with the discussion in section \ref{sect:lcs}. After a brief description of the method we used to find the minima, we present two examples of sequences of minima.
\begin{figure}[tb]
\centering
\includegraphics[height=7cm]{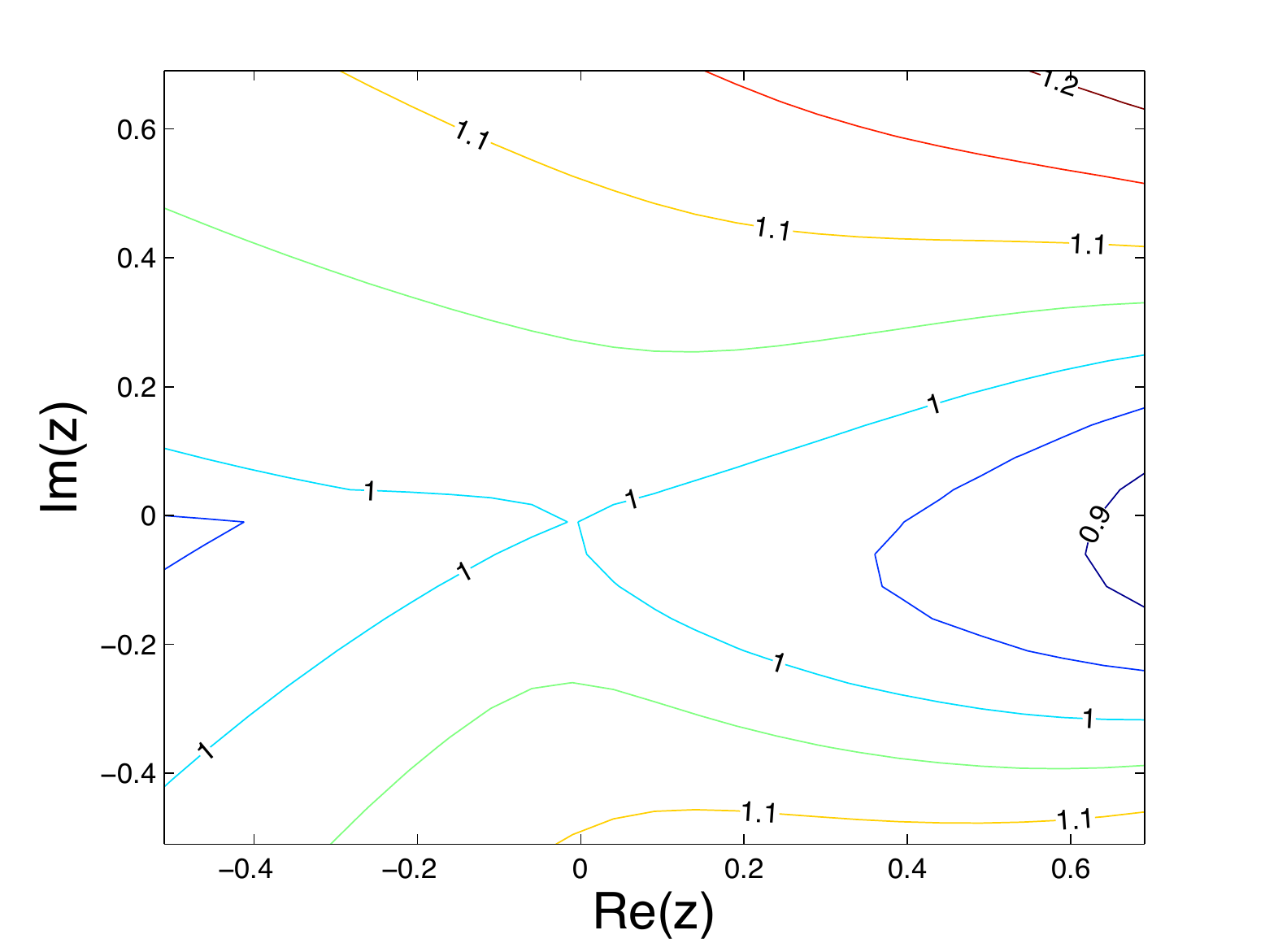}
\caption{The ratio of the scalar potential computed with the Meijer functions and the potential computed using the LCS expansions, for the same fluxes as the previous figure. The closer to 
the LCS point, the better the match between the two potentials.}
\label{fig:pot_comp}
\end{figure}

To speed up the numerical calculation of the potential, we proceed as follows. We first compute the periods and their derivatives on a grid in the complex structure modulus plane. This computation is performed using the built-in Meijer functions of Maple for the full periods, and using Matlab for the LCS expansions of the periods. We then feed these periods into Matlab where the superpotential, K\"ahler and scalar potentials are computed. We also use Matlab to find the minima of the potential, and determine their position and minimum value of the potential. 
\begin{figure}[htb]
\centering
\includegraphics[height=9cm]{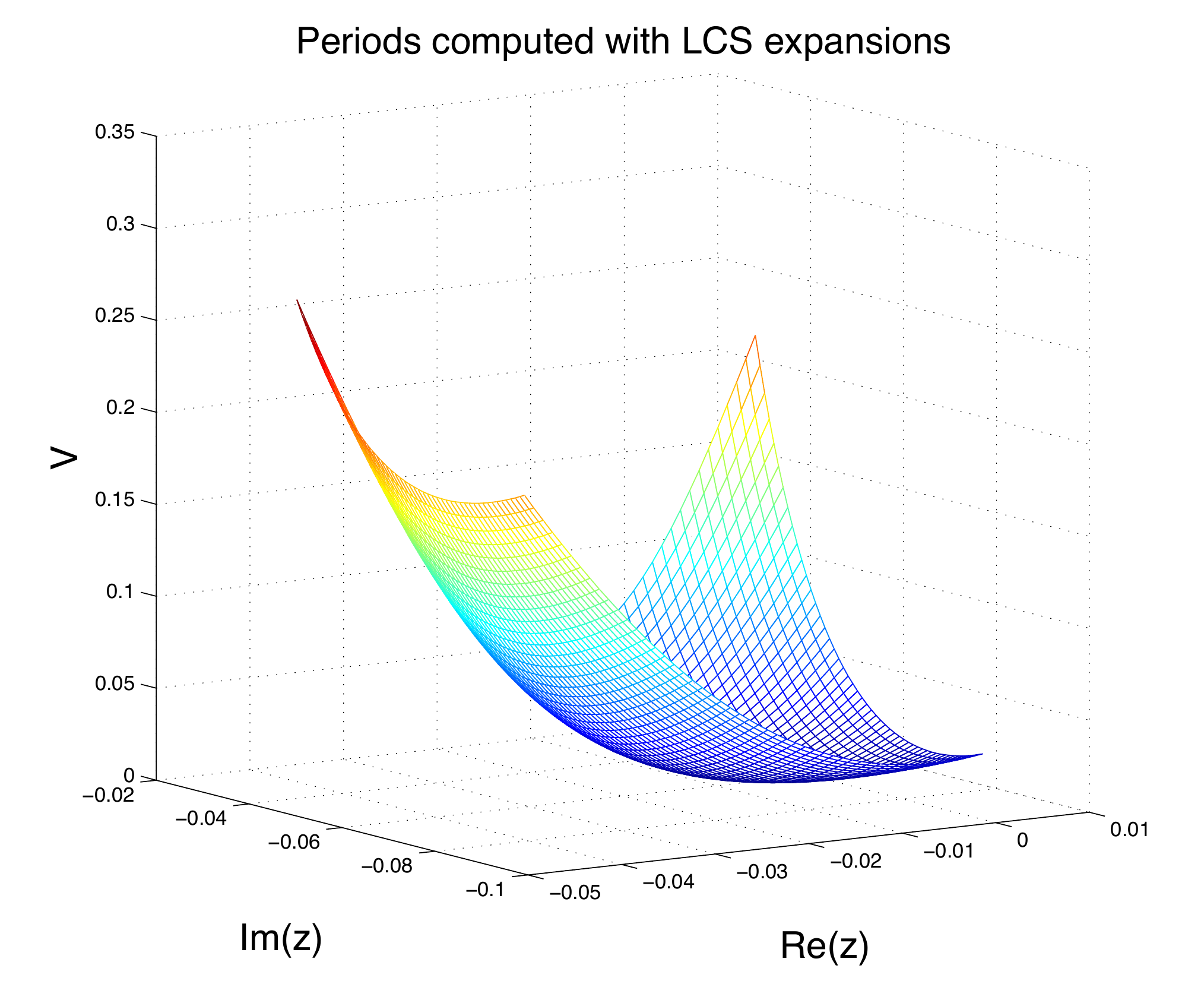}
\caption{Using the LCS expansions of the periods allow us to zoom in on the supersymmetric Mirror Quintic minimum of figure \ref{fig:meijer_lcs}.}
\label{fig:lcszoom}
\end{figure}

Since the minima in the series approach the LCS point, the LCS expansion of the periods provides a good and computationally cheap approximation of the features of the minima closest to this point. An illustration of this is shown in the figures \ref{fig:meijer_lcs} and \ref{fig:pot_comp}, where the Mirror Quintic potential for the flux configuration $H=(-2,-4,-33,0)$ and  $F = (3,-18,9,-1)$ is plotted using both the full Meijer functions and the LCS expansions. As can be seen from the figures, the two potentials are very similar; in particular the location and value of the potential in the minimum agree to a good degree. Consequently, the LCS expansions determine the features of minima to a good approximation at least for $|z| < 0.2$.

Given that the periods are computed on a grid, the position of a minimum of the potential can never be determined to a better accuracy than the grid spacing. Thus minima that lie closer to the LCS point remain undetected until the grid spacing is refined. For computationally expensive functions such as the Meijer functions, this provides a significant obstacle, in that refining the grid soon becomes practically impossible. On the other hand, the LCS expansions are simple functions that can easily be computed on more and more refined grids. In figure \ref{fig:lcszoom} we show a more detailed picture of the Mirror Quintic minimum that was obtained using the LCS expansions of the periods. 

Thus, in order to investigate whether the series of minima reported on in \cite{Ahlqvist:2010ki} continue indefinitely, we use the LCS expansion of the periods. We first compute the potential for a flux configuration on a sparse grid, identifying the region in the $z$-plane where the minimum is located. At this stage, we also note if we need to move the branch cut that emerges from the LCS point in order to trace the minimum to another level in the potential.\footnote{In some cases, it is necessary to move several steps down in the potential spiral to find the minimum, and for some flux values no minimum is found, even at the lowest level of the potential.} We then zoom in on the region that should contain a minimum and recompute the potential on a narrow grid around this point. This allows us to compute the location and potential value of the minimum to a higher accuracy. We then act on the flux vectors with the conifold monodromy matrices, and repeat the calculations for the next minimum in the series. 
\begin{figure}[tb]
\centering
\includegraphics[height=8cm]{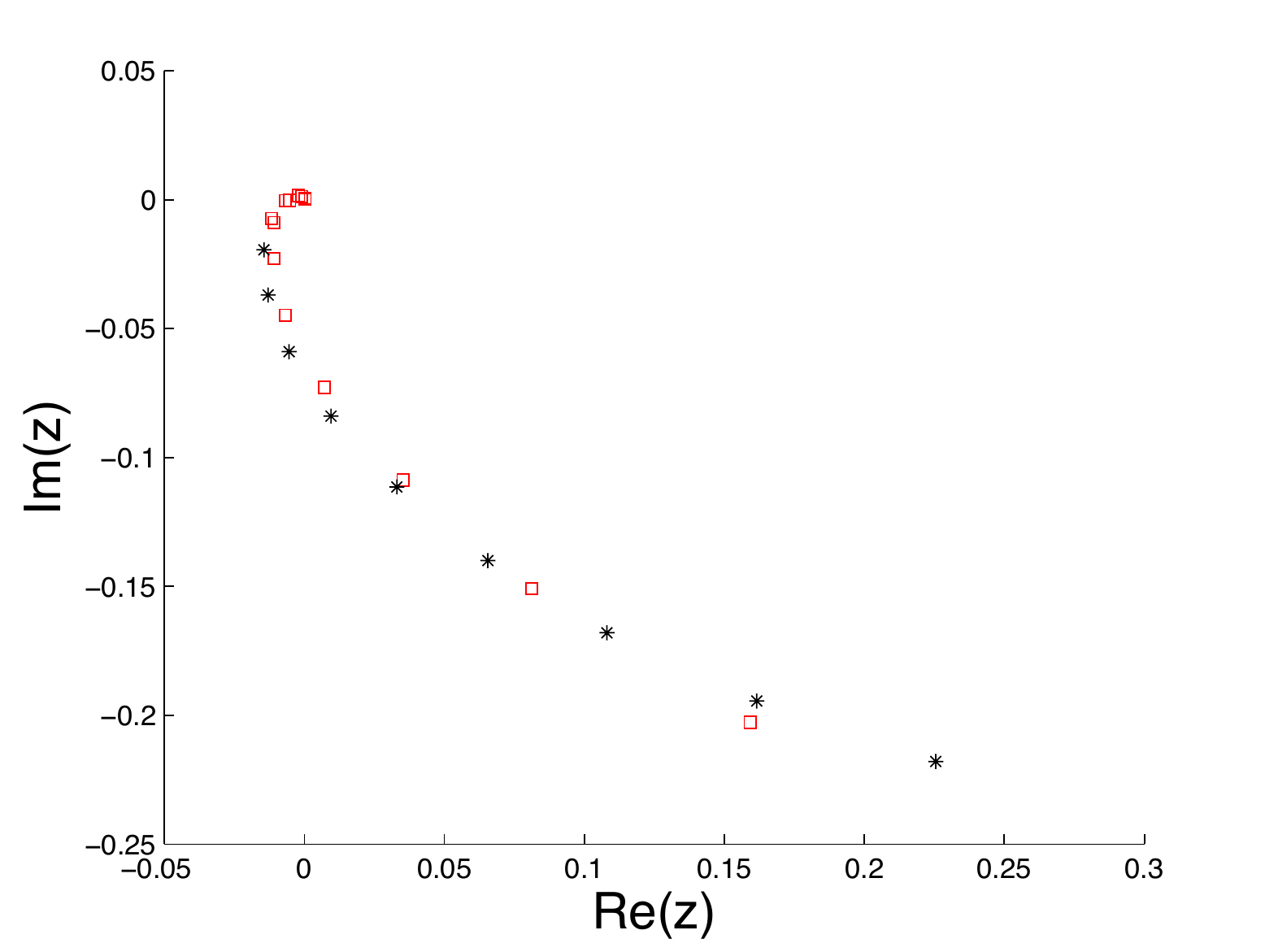}
\caption{The distribution in the $z$ plane of a series of minima that approach the large complex structure point in the Mirror Quintic moduli space. The minima have NS-NS flux $H=(-2,-4,-33,0)$ and RR flux $F = (F^0,-18,9,-1)$, where $F^0$ ranges from -17 to 9. The red squares indicate minima with negative $F^0$, whereas black stars are used for minima with positive $F^0$.}
\label{fig:zseries}
\end{figure}
\begin{figure}[htb]
\centering
\subfigure{\includegraphics[height=5cm]{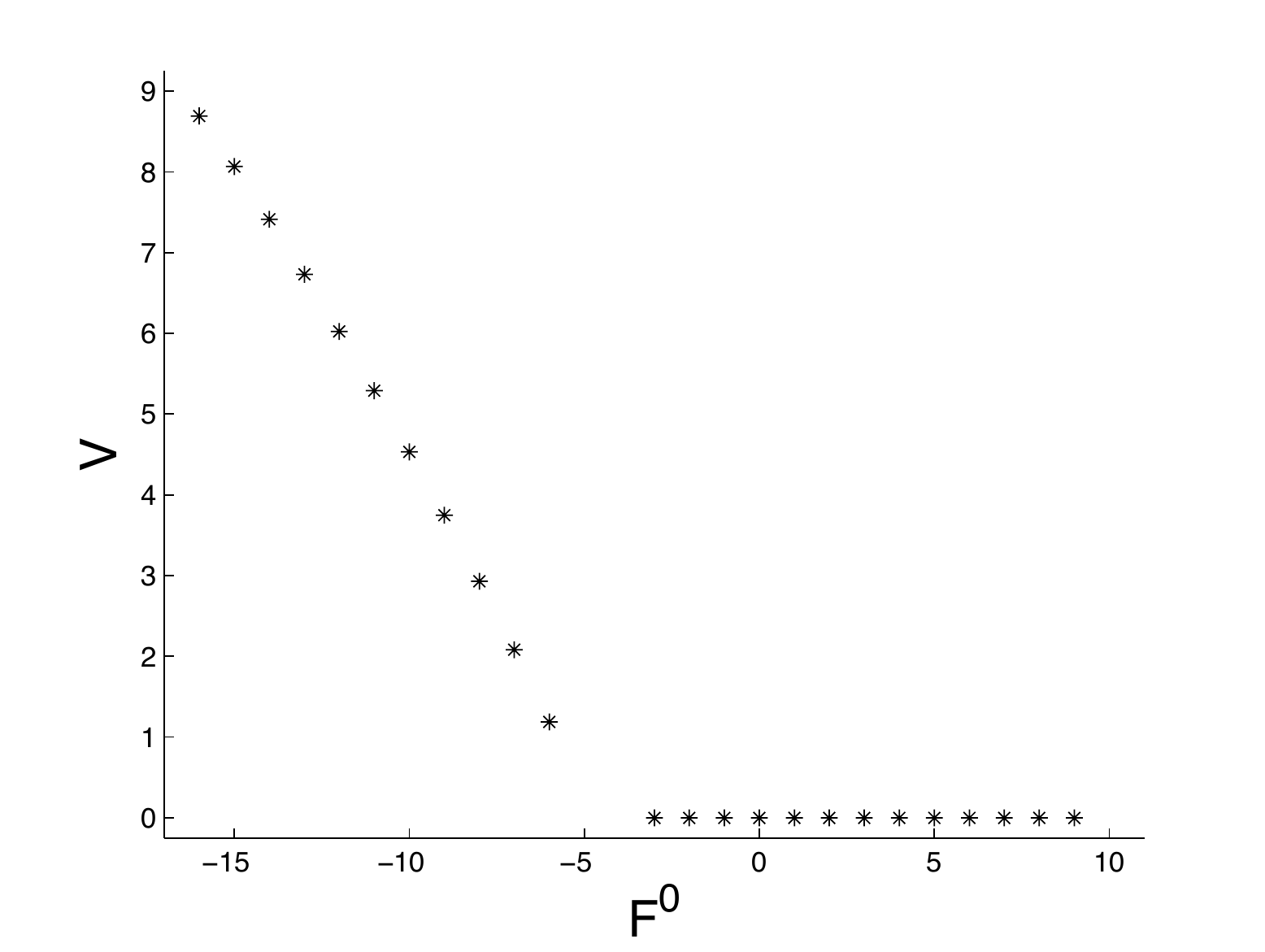}}
\qquad
\subfigure{\includegraphics[height=5cm]{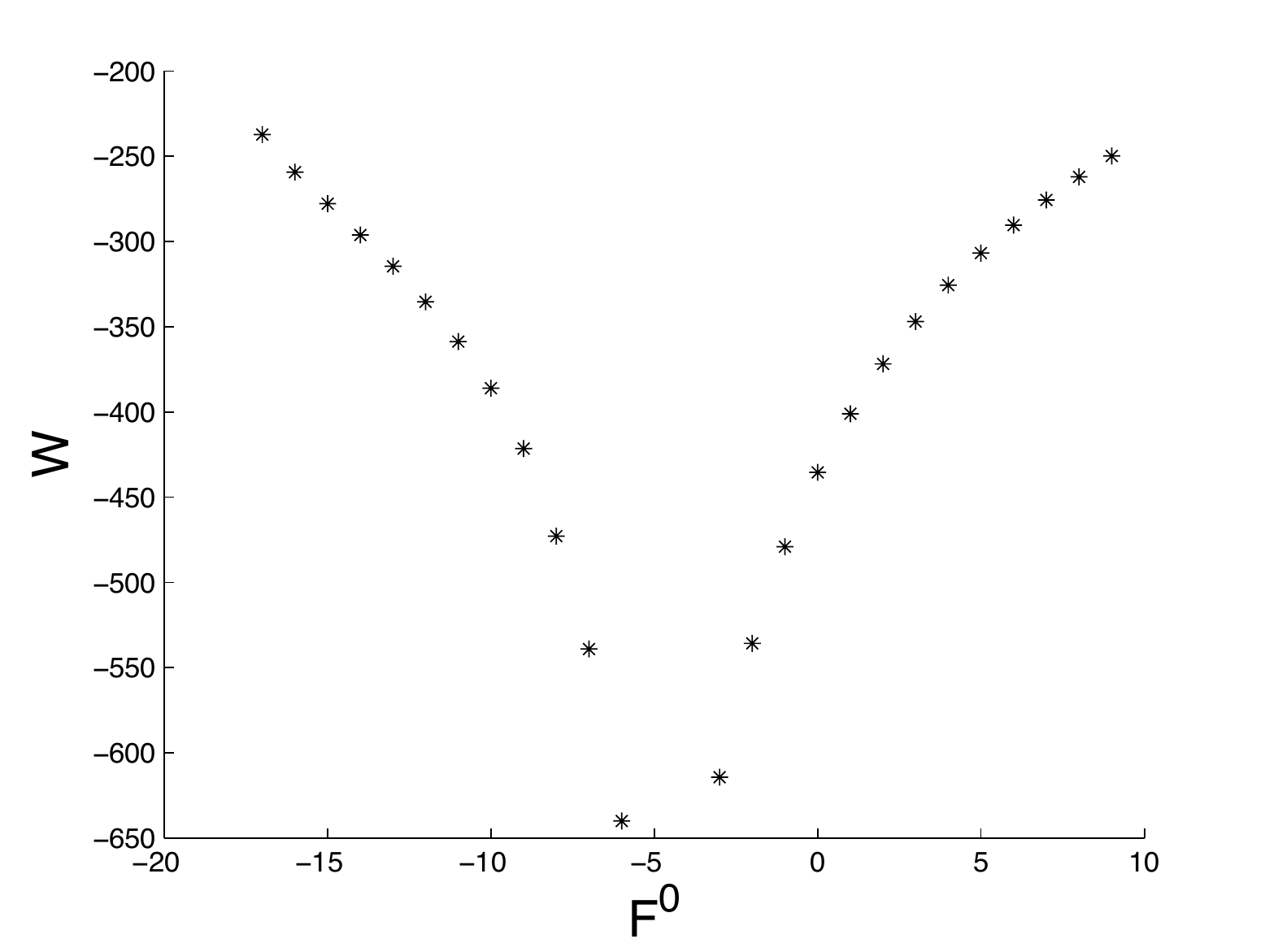}}
\qquad
\subfigure{\includegraphics[height=5cm]{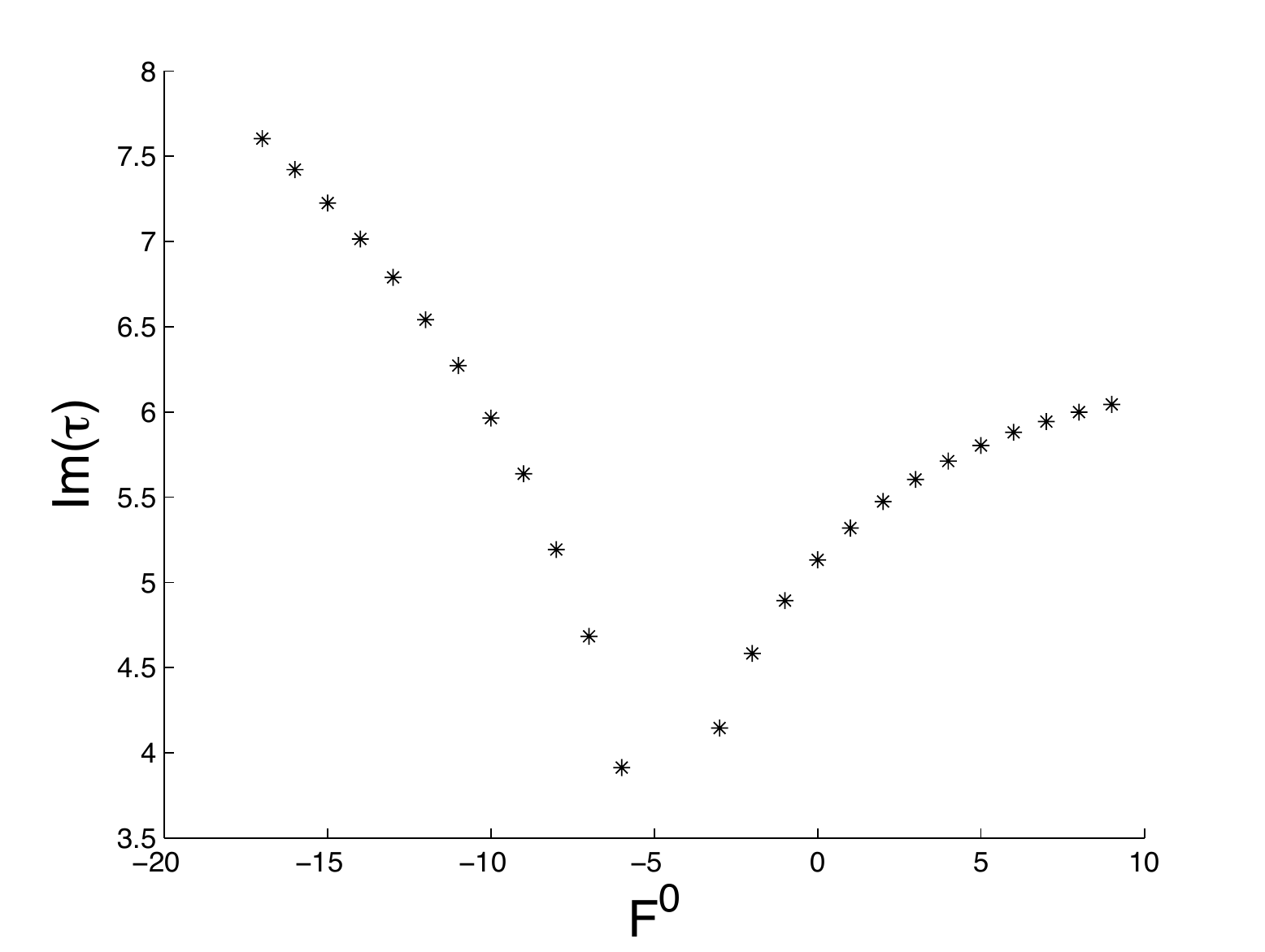}}
\caption{Here we show the value of the potential, the superpotential and the imaginary part of the axio-dilaton $\tau$ for the series of Mirror Quintic minima with NSNS flux $H=(-2,-4,-33,0)$ and RR flux $F = (F^0,-18,9,-1)$. As can be seen from the first panel, all minima with positive $F^0$ are ISD and have vanishing scalar potential. The value of the superpotential is large and negative for all minima in the series, and the dilaton does not run away to zero or infinity.}
\label{fig:VWseries}
\end{figure}

\subsection*{A series of minima on the Mirror Quintic}

Using the outlined procedure, we reproduce the minima with $F^0 = 3 ... 9$ in the Mirror Quintic series reported on in table 3 and figure 5 of \cite{Ahlqvist:2010ki}. In addition we find new minima with $F^0 = -17 ... -6$ and $-3 ... 2$. We found no minima for the two values $F^0 = -5, -4$, despite having studied the downward spiral of the scalar potential until it reaches its lowest level and turns back up. The $z$-distribution of the minima in the series is shown in figure \ref{fig:zseries}. As can be seen, starting from $F^0=9$ the series of minima approaches the LCS point for decreasing values of $F^0$. However, as the by now negative $F^0$ increases in magnitude, the minima again recede from the LCS point, until they leave the region where the LCS expansion can be trusted. Thus, this series is not infinite.

As shown in figure \ref{fig:VWseries}, all minima with positive $F^0$ have vanishing potential in the minimum, and fulfill the ISD condition. Conversely, the minima with negative $F^0$ have a non-zero potential value. Thereby, this example confirms our general result that the series of minima that converge to the LCS point eventually break the ISD condition, thus inducing non-zero F-terms also in the complex structure and axio-dilaton directions. 

Figure \ref{fig:VWseries} also shows the vacuum expectation value for the superpotential for the series of minima. Since this is large for all minima, supersymmetry is broken by the K\"ahler moduli, which have non-zero F-terms. We note that the tadpole for this series of minima is high, so the phenomenological interest of these minima is fairly limited. 

From figure \ref{fig:VWseries} we also see that $\mbox{Im} \tau$ does not run away, but stays in the range $4-8$. Consequently, $\cG_\tau$ does not degenerate, and therefore this series of minima does not lie in a decompactification limit of the axio-dilaton part of moduli space.

\subsection*{A series of minima on Model 12}
\begin{figure}[tb]
\centering
\includegraphics[height=9cm]{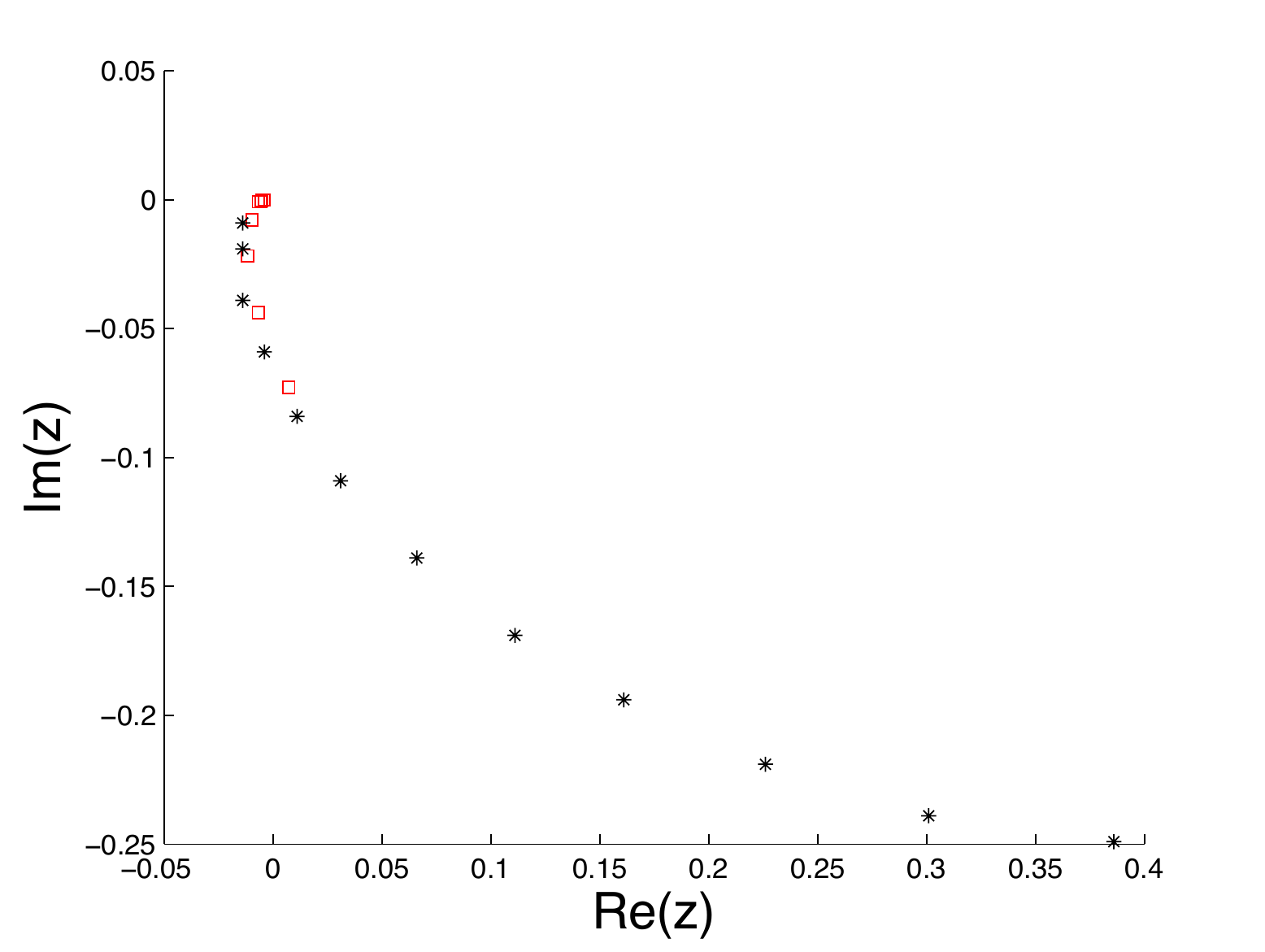}
\caption{The distribution in the $z$ plane of a series of minima that approach the large complex structure point in the moduli space of Model 12. The vacua have NSNS flux $H=(-2,-4,-33,0)$ and RR flux $F = (F^0,-18,9,-1)$, where $F^0=-12...11$. The red squares indicate minima with negative $F^0$, whereas black stars are used for minima with positive $F^0$.}
\label{fig:zseries12}
\end{figure}

The longest series of minima that was reported on in table 3 and figure 5 of \cite{Ahlqvist:2010ki} was found on the one-parameter Calabi--Yau known as Model 12. This series consists of twenty-nine minima, with NS-NS flux $H=(-2,-4,-33,0)$ and RR flux $F = (F^0,-18,9,-1)$, $F^0 = 7, ... ,36$. Using the LCS expansions of the periods, we reproduce some minima of this series and extend it to smaller values of $F^0$, as shown in figure \ref{fig:zseries12}. Just as for the Mirror Quintic example, we find that more minima exist in the vicinity of the LCS point, but the minima bounce out from the LCS point again as $F^0$ becomes large and negative. Thus, this series of minima does not continue indefinitely.

\begin{figure}[tb]
\centering
\subfigure{\includegraphics[height=5cm]{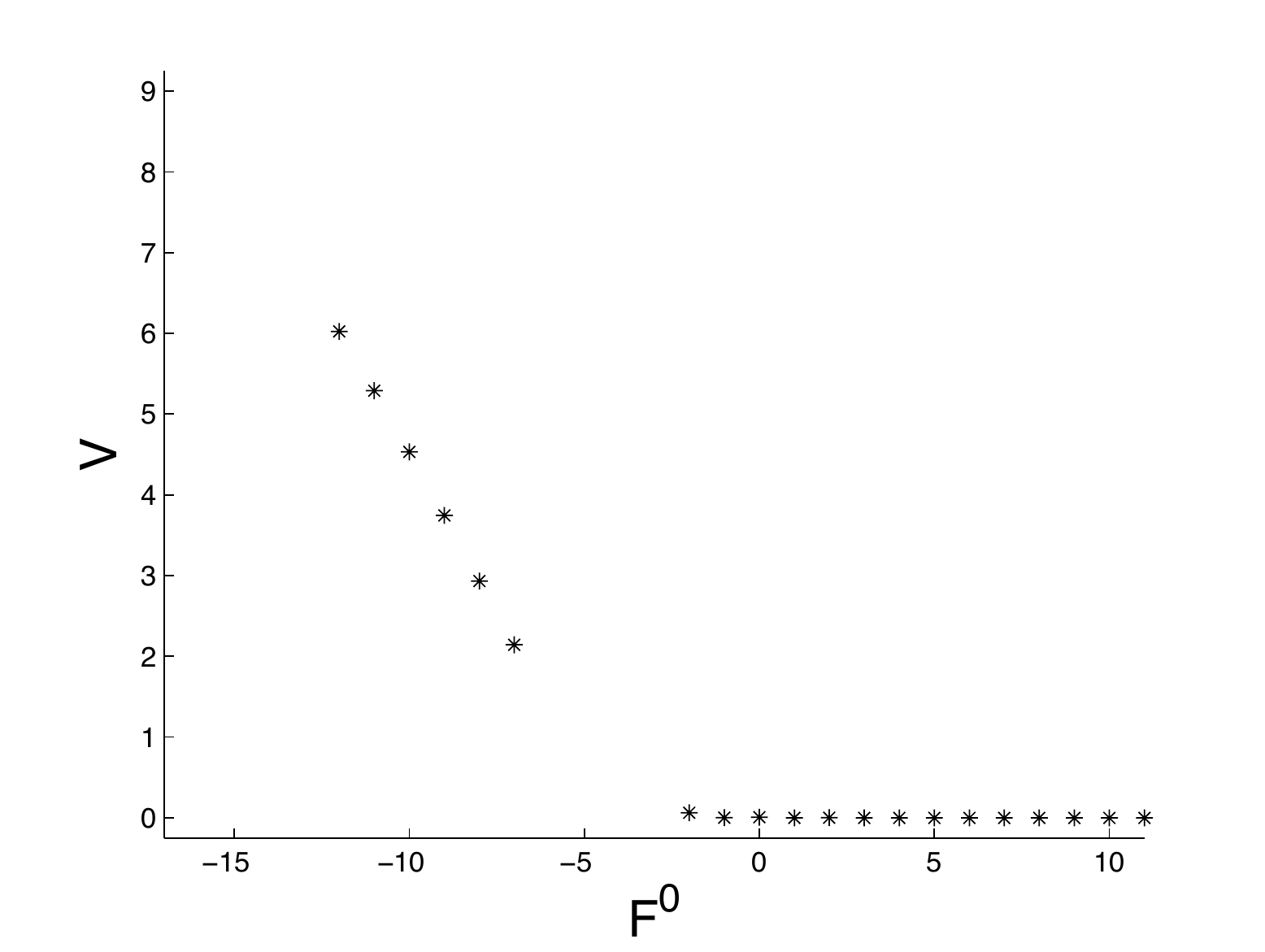}}
\qquad
\subfigure{\includegraphics[height=5cm]{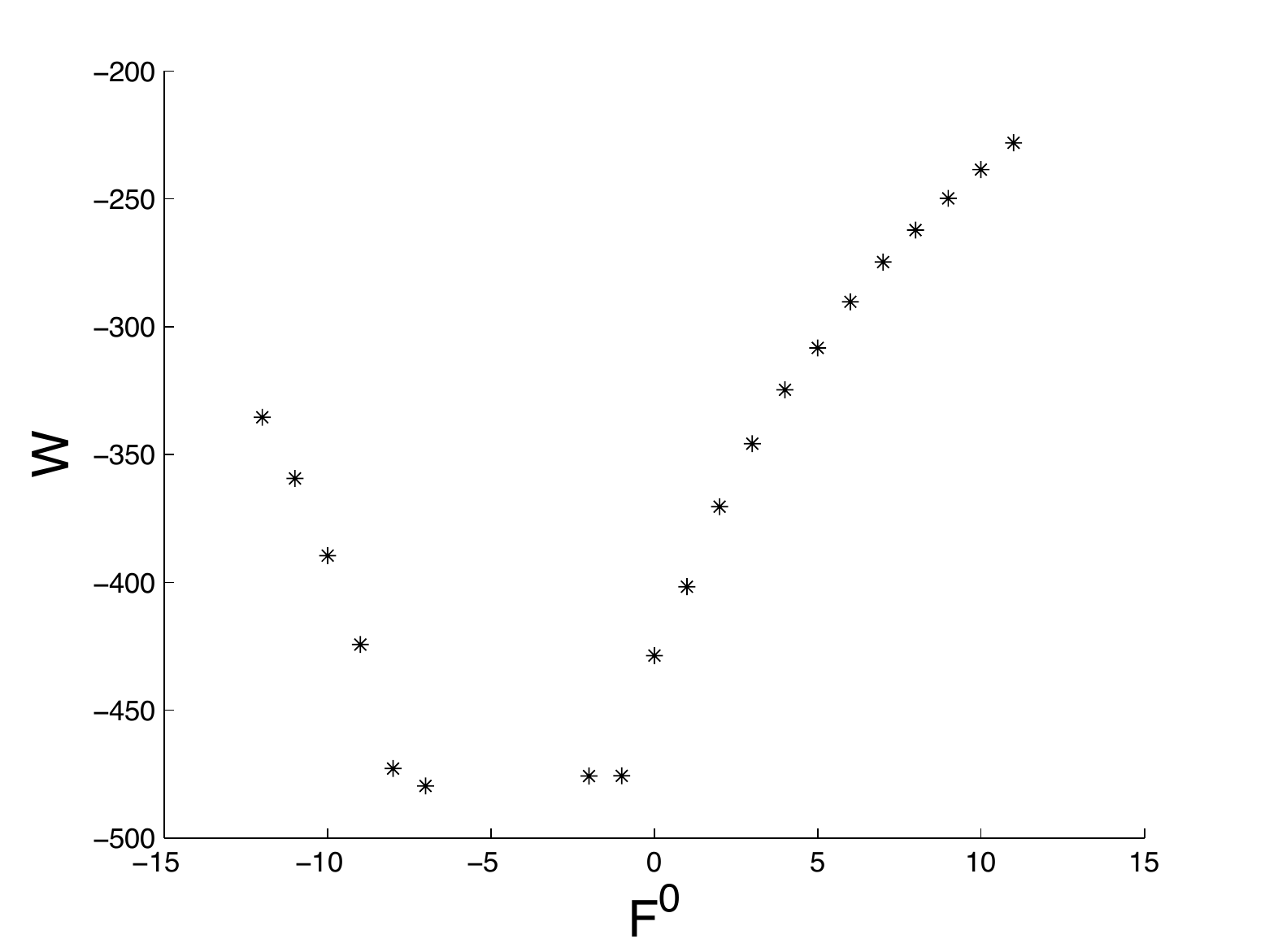}}
\qquad
\subfigure{\includegraphics[height=5cm]{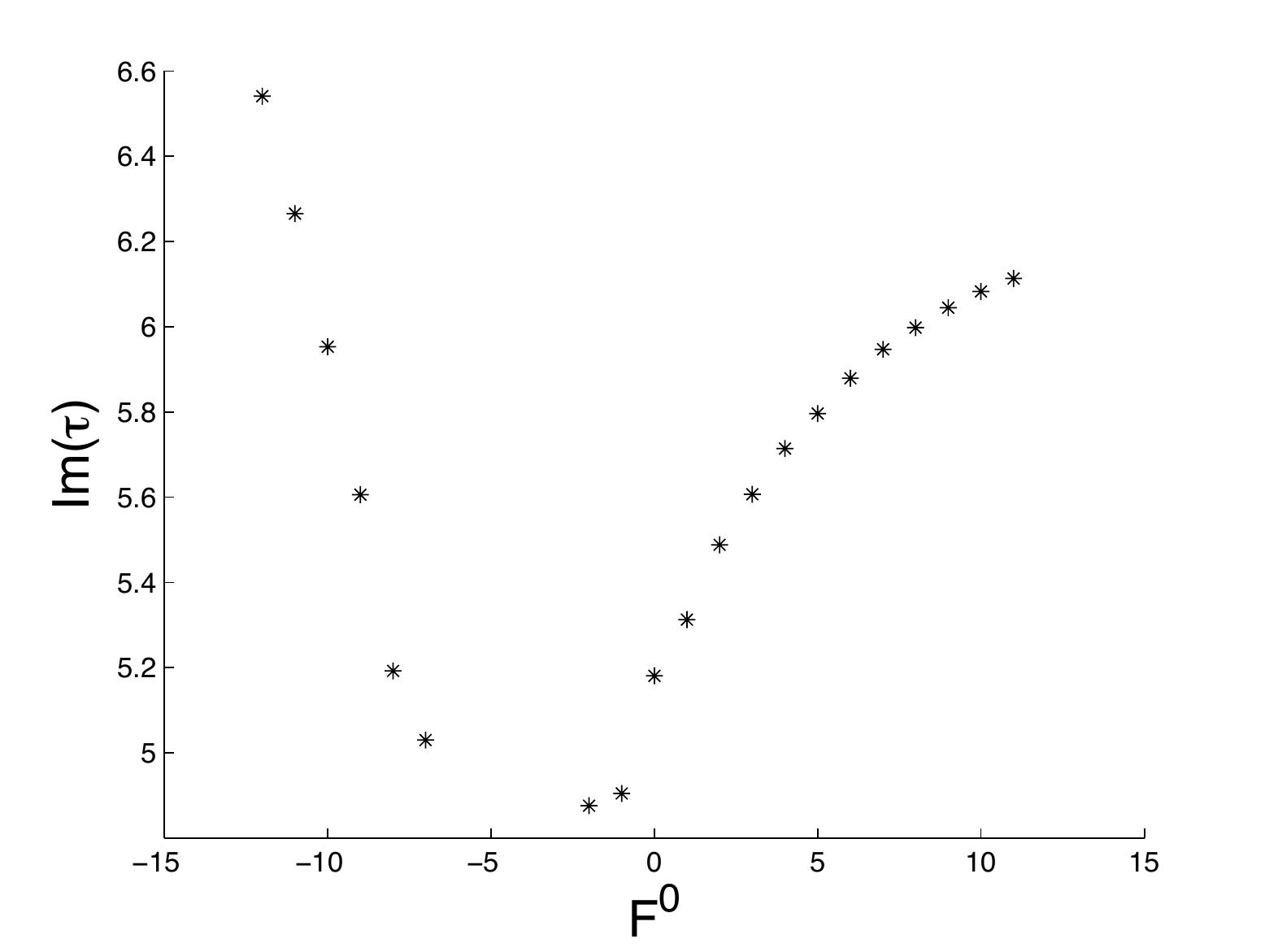}}
\caption{The value of the potential, the superpotential and the imaginary part of the axio-dilaton $\tau$ for the series of Model 12 minima with NSNS flux $H=(-2,-4,-33,0)$ and RR flux $F = (F^0,-18,9,-1)$ of Model 12. The features of this series of minima closely parallels those of the Mirror Quintic series.}
\label{fig:VWseries12}
\end{figure}

The value of the potential, superpotential and $\mbox{Im} \tau$ for Model 12 are presented in figure \ref{fig:VWseries12}. As can be seen, the features are similar to the Mirror Quintic series. As expected, the ISD condition is eventually broken for negative values of $F^0$, and $\mbox{Im} \tau$ stays finite for the whole series. The superpotential is large and negative also for this series, and the tadpole is the same as for the Mirror Quintic series.

\section{Conclusions and outlook}

In this paper we have extended the no-go result of Ashok and Douglas to include also regions around certain D-limits.
For a class of one-parameter models we studied the large complex structure limit, the conifold point and the decoupling limit,
and found that none of these can support infinite sequences of ISD vacua. This analysis was performed by explicitly computing 
a certain positive definite quadratic form defined on the space of flux quanta. This form gives the total D3-brane charge originating from
three-form flux in the case of an ISD vacuum. By analysing the precise form of the eigenvalues and eigenvectors as the various D-limits are
approached we demonstrated that no infinite sequences are possible. We also extended this analysis to the LCS limit of a two-parameter model, again finding that no infinite sequences exist. Furthermore, we explained how infinite sequences accumulating to D-limits in $K3\times K3$ compactifications really correspond to finitely many vacua after the automorphism group is taken into account.

To complement the analytical results, we studied two of the sequences found by Ahlqvist et al. \cite{Ahlqvist:2010ki} numerically.
We used expansions around the LCS point to facilitate the computations of the periods, thus making a fine grid possible. 
The sequences were found to turn close to the LCS point and then be repelled from it, eventually violating the ISD condition, perfectly 
in line with the analytical results.

In the present work we used fairly pedestrian methods to analyse the structure of the quadratic form around various singularities. For this, we needed expansions 
of the periods in the D-limit under consideration. This is in contrast to the statistical analysis, where the number of vacua is estimated without such detailed 
understanding of the Calabi--Yau. Although our method requires more information, it allows us to refine the results of the statistical analysis in the models we consider.
It would of course be very interesting to formulate more general and transparent conditions on the singularity required for infinite sequences. Such a result would be a step 
towards a more general finiteness theorem.

Additionally, two interesting directions of future research would be to investigate whether similar techniques 
can be applied also in the case of generalized Calabi-Yau manifolds and to analyse how warping corrections 
affect the results for sequences accumulating to a conifold point.

\acknowledgments We thank Ralph Blumenhagen, Michael R. Douglas, Matthew C. Johnson, Dieter L\"ust and Gonzalo Torroba for discussion and comments. The work of APB is supported by the Austrian Science Foundation (FWF) under grant I192.
The research of NJ is supported by the START project Y435-N16 of the FWF and by the FWF project P21927-N16.  
ML's research is supported by the Munich Excellence Cluster for Fundamental Physics ``Origin and the Structure of the Universe''. 
The work of N-OW is supported by the FWF under grants P21239 and I192.

\appendix

\section{Expansions around LCS points}\label{app1}

\subsection{One-parameter models}

For a one-parameter model, the period vector takes the following general form around the LCS point \cite{Ahlqvist:2010ki}
\eq{
\begin{pmatrix} \Pi_3 \\ \Pi_2 \\ \Pi_1 \\ \Pi_0  \end{pmatrix} \sim \begin{pmatrix} \al_3 \, t^3 + \ga_3 \, t + i\de_3 
\\ \be_2 \, t^2 + \ga_2 \, t + \de_2 \\ t \\ 1
\end{pmatrix}.
}{eq:app1} 
Here $t \sim -i\log z$, and the LCS point is at $\im t \to \infty$. All coefficients except $\de_3$ are rational.
For the models we study, the coefficients are presented in table \ref{tb:1}.

Let $t = t_1 + it_2$ with $t_{1,2}\in \bR$. For general expansion coefficients we then get the following expansions around $t_2 = \infty$
\eq{
e^{-K} = (-2\be_2-2\al_3)t_2^3 + (2\de_2 + 6 \al_3 t_1^2-2\be_2 t_1^2+2 \ga_3) t_2 + \ldots \ ,
}{eq:ap2}
\eq{
\cG_t = \begin{pmatrix} g_{11} \, t_2^3 + \cO(t_2) & g_{12} \, t_2 + \cO(1/t_2) & g_{13}\,t_2 + \cO(1/t_2) & g_{14}\frac{1}{t_2} + \cO(1/t_2^3) \\
\cdot & g_{22}t_2 + \cO(1/t_2) & g_{23}\frac{1}{t_2} + \cO(1/t_2^3) & g_{24}\frac{1}{t_2} + \cO(1/t_2^3)\\
\cdot & \cdot & g_{33}\frac{1}{t_2} + \cO(1/t_2^3) & g_{34}\frac{1}{t_2^3} + \cO(1/t_2^5)\\
\cdot &\cdot & \cdot  & g_{44}\frac{1}{t_2^3} + \cO(1/t_2^5)
        \end{pmatrix} \ .
}{eq:ap3}
The coefficients $g_{ij}$ are a little messy:
\begin{align}
&g_{11} = -\frac{2\al_3^2 \, (9\al_3+5\be_2)}{(\al_3+\be_2) (9 \al_3+\be_2)} &g_{12} &= -\frac{2\, \al_3 \big[(\be_2-3\al_3)\ga_2+3 \be_2(5 \al_3 + \be_2) t_1 \big] }{(\al_3+ \be_2) (9\al_3+ \be_2)}\\
&g_{13} =\frac{2\al_3 \, (3\al_3-\be_2) }{(\al_3+\be_2)(9\al_3+\be_2)}  &g_{22} &= -\frac{2\be_2^2 \, (5\al_3+\be_2)}{(\al_3+\be_2)(9\al_3+\be_2)}\\
&g_{14} = 9\, t_1 \, g_{13}   &g_{23} &= -\frac{2\, \big[\ga_2(5 \al_3 +\be_2)+\be_2(\be_2 + 13 \al_3) t_1\big]}{(\al_3+\be_2)(9\al_3+\be_2)}\\
&g_{24} = \frac{\be_2}{\al_3}g_{13} &g_{33} &= \frac{g_ {22}}{\be_2^2} \\
&g_{34} = 3 t_1 g_{33} &g_{44} &= \frac{g_{11}}{\al_3^2}\, . &
\end{align}
Note that special relations among the coefficients can change the asymptotic behaviour. E.g., for all models in \cite{Ahlqvist:2010ki} we have
\eq{
\be_2 = 3\al_3
}{eq:ap5}
yielding
\eq{
g_{13} = g_{14} = g_{24} = 0.
}{eq:ap6}
Specifically, for the mirror quintic values, the expansion of $\cG_t$ is
\eq{
\cG_t = \begin{pmatrix} \frac{5}{6} \, t_2^3 + \cO(t_2) & \frac{5t_1}{2} \, t_2 + \cO(t_2^{-1}) & -\left(\frac{5}{6} + t_1^2\right)\,\frac{1}{t_2} + \cO(t_2^{-3}) & \frac{-10 t_1^3-25t_1+ 12 i \de_3}{10}\frac{1}{t_2^3} + \cO(e^{-t_2}) \\
\cdot & \frac{5}{2}t_2 + \cO(t_2^{-1}) & -\frac{10 t_1 +11}{5}\frac{1}{t_2} + \cO(t_2^{-3}) & \frac{-30 t_1^2-66 t_1+25}{10}\frac{1}{t_2^3} + \cO(t_2^{-5})\\
\cdot & \cdot & \frac{2}{5}\frac{1}{t_2} + \cO(t_2^{-3}) & \frac{6\, t_1}{5}\frac{1}{t_2^3} + \cO(t_2^{-5})\\
\cdot &\cdot & \cdot  & \frac{6}{5}\frac{1}{t_2^3} + \cO(t_2^{-5})
        \end{pmatrix} \ .
}{eq:ap7}
The K\"ahler covariant derivative of the period vector has the expansion
\eq{
\begin{pmatrix} D_t\Pi_3 \\ D_t\Pi_2 \\ D_t\Pi_1 \\ D_t\Pi_0  \end{pmatrix} \sim \begin{pmatrix} A_3 \, t_2^2 + B_3 \, t_2 + C_3 + \ldots 
\\ B_2 t_2 + C_2 + \ldots \\ C_1  + \frac{D_1}{t_2} + \ldots \\ \frac{D_0}{t_2} + \frac{E_0}{t_2^2} + \ldots
\end{pmatrix}, 
}{eq:ap8} 
where
\begin{align}
A_3 &= -\frac{3}{2}\al_3 & B_3 &= -\frac{i \al_3 t_1 (3\al_3 - 5 \be_2)}{2(\al_3 + \be_2)}  &B_2&=\frac{i\be_2}{2}  
&C_2&=-\frac{4\al_3 \be_2 t_1}{\al_3 + \be_2} - \frac{\ga_2}{2} \\
C_1 &= -\frac{1}{2}& D_1 &= \frac{i t_1 (9\al_3 + \be_2)}{2(\al_3 + \be_2)} & D_0&=\frac{3i}{2} & E_0 &= \frac{t_1(3\al_3 - \be_2)}{\al_3 + \be_2}.
\end{align}

\begin{table}
\begin{center}

\begin{tabular}[tb]{ |l | c | c | c || c | c| c | }
  \hline 
  Model & $\al_3$ & $\ga_3$ & $\de_3$ & $\be_2$ & $\ga_2$ & $\de_2$\\
\hline
  Mirror Quintic: & $ -\frac{5}{6}$ & $-\frac{25}{12}$ & $\frac{200 \zeta(3)}{(2\pi)^3}$ & $-\frac{5}{2}$ & $-\frac{11}{2}$ & $\frac{25}{12}$ \\
\hline  
Model 12: & $-\frac{2}{3}$ & $-\frac{5}{3}$ & $\frac{18 \zeta(3)}{(\pi)^3}$ & $-2$ & $-5$ & $\frac{5}{3}$ \\
  \hline  
\end{tabular}
\caption{Expansion coefficients around the LCS points for the considered one-parameter models.}\label{tb:1}
\end{center}
\end{table}

\subsection{Coefficients of the metric $\cG_z$ of the two--parameter model} \label{App2para}
The expansion of the metric of the complex structure moduli space of the model $\mathcal{M}_{(86,2)}$ near to the LCS is given in formula (\ref{eqn:Gz_twoparameter}). Here we list its coefficients $a_{ij}$: 
\begin{align}
a_{11} &= \frac{17}{6} \nonumber
&a_{12} &= -\frac{545}{864}\,x_1-\frac{61477}{10368}\, y_1 \nonumber\\ 
a_{13} &=-\frac{109 }{72}\,x_1-\frac{545}{864} \,y_1  
&a_{14} &= -\frac{109}{144}\,x_1y_1-\frac{545}{1728}\,y_1^2-\frac{545}{1728}\nonumber \\ 
a_{15}&= \frac{545}{1728}\, x_1y_1-\frac{26651}{20736}\,y_1^2-\frac{11963}{20736} 
&a_{16} &= -\frac{109}{144}x_1y_1^2-\frac{109}{288}\,x_1-\frac{545}{1728}y_1^3-\frac{2725}{3456}\,y_1\nonumber\\ 
a_{22} &= \frac{61477}{10368}
&a_{23} &=\frac{545}{864}\nonumber\\
a_{24} &=\frac{109}{144}\,x_1+\frac{545}{864}\,y_1+\frac{109}{1152}
&a_{25}&= -\frac{545}{1728}\,x_1+\frac{26651}{10368} \,y1-\frac{545}{13824}\nonumber \\
a_{26} &=\frac{545 }{576}y_1^2+\frac{109 }{72}x_1 y_1+\frac{109}{576} y_1-\frac{2725}{3456}
&a_{33}&=\frac{109}{72}\nonumber\\
a_{34} &=\frac{109}{144}\,y_1
&a_{35}&=-\frac{545}{1728}\,y_1\nonumber\\
a_{36} &=\frac{109}{144}\,y_1^2-\frac{109}{288}
&a_{44}&=\frac{109}{576}\nonumber\\
a_{45} &=-\frac{545}{6912}
&a_{46}&=\frac{109}{288}\,y_1\nonumber\\
a_{55} &=\frac{61477}{82944}
&a_{56}&=\frac{109}{288}\,x_1\nonumber\\
a_{66} &=\frac{109}{288} \, .
\end{align}

\bibliographystyle{fullsort}

\end{document}